\documentclass[aps,prx,reprint,groupedaddress]{revtex4-2}

\usepackage{mathrsfs}
\usepackage{array}
\usepackage{amsmath}
\usepackage{amssymb}
\usepackage{graphicx}
\usepackage{color,xcolor}
\usepackage[dvipsnames]{xcolor}
\usepackage{booktabs}
\usepackage{amstext}
\DeclareMathOperator{\Tr}{Tr}
\usepackage{amsfonts}
\usepackage{float}
\usepackage{bm}
\usepackage{epstopdf}
\usepackage{color}
\usepackage{subfigure}
\usepackage{hyperref}
\usepackage{calrsfs}
\usepackage{mathtools}

\begin{document}


\title{Active quantum memory: Exploring the quantum dynamical process of voltage-gated ion channel}


\author{Yu-Juan Sun}
\email{sunyj28@phys.ncku.edu.tw}
\affiliation{Department of Physics and Center for Quantum Information Science, National Cheng Kung University, Tainan 70101, Taiwan}
\author{Wei-Min Zhang}
\email{wzhang@mail.ncku.edu.tw}
\affiliation{Department of Physics and Center for Quantum Information Science, National Cheng Kung University, Tainan 70101, Taiwan}


\date{\today}

\begin{abstract}
It is known that the opening or closing mechanism of a voltage-gated ion channel is triggered by the potential difference across the cell membrane in the nervous system. Based on this picture, we model the ion channel as a nanoscale two-terminal ionic tunneling junction. 
We apply an external time-varying voltage on the junction to mimic the membrane potential difference in the stimulation of neurons. We derive the non-Markovian quantum Langevin equation from the first principle of quantum mechanics for the ion transport in ion channels, and obtain the ion transport current in terms of quantum tunnelings of ions controlled by the time-varying voltage. 
We find that the time-varying voltage induces an effective magnetic flux, which causes quantum coherence in ion tunnelings. This effective magnetic flux induces further an oscillatory component to the spectral structure of the ion system, forming a time-dependent quantum memory. Such voltage-induced memory is defined as the active quantum memory, which manifests in the system with a regular oscillatory sideband structure in the ion transport current. 
The sideband structure demonstrates a multi-crossing hysteresis in the I-V curve, responding to the variation of the time-varying voltage (membrane potential difference). We also find that the strength of active quantum memory can be described by the ratio of amplitude and frequency of the time-varying voltage, and can be quantitatively measured through the number of non-zero cross points in the current-voltage (I-V) hysteresis and conductance-voltage (G-V) diagram. 
Additionally, we explore the temperature dependence of the active quantum memory in such a system. Further, we apply this model to the ion channel system on the biological energy scale. The description of these active quantum memory characteristics provides a quantum mechanical understanding to the underlying mechanism of ion channel dynamics.
\end{abstract}


\maketitle

\section{Introduction}

It is widely accepted that nerve status can be demonstrated by the electrical signals carried via the potential difference between the intracellular and extracellular ion solutions of the cell membrane. The corresponding ion channel responses dominate the overall changes. This mechanism makes the nervous system look like a circuit system. Taking the voltage-gated ion channel, for example, there are plenty of ion channels on the neuron membrane. The voltage-gated ion channel can be taken as a nonlinear classical circuit in electrophysiology experiments. From the electrophysiological perspective, researchers have focused on establishing a practically useful circuit theory to describe the behavior of ion channels. The prototypical example is the HH model, proposed by Hodgkin and Huxley \cite{HH1952}, utilizing the nonlinear RC circuit. Its extensions have successfully depicted the conductance of various ion channel clusters observed in experiments. However, while the HH model described the macroscopic channel conductance successfully, the nonlinear 
classical resistance-capacitance (RC) circuit description does not capture the microscopic physics of individual ion channels. Moreover, when the HH model was just proposed, experimental recordings of single ion channels were not yet available. Such measurements became possible only after the development of the patch clamp method by Neher and Sakmann \cite{NS1976,HMNSS1981}. Although the patch clamp method enabled the direct observation of single ion channel activity and revealed the stochasticity of channel opening or closing, the underlying mechanism of ion permeation remained unclear.

In fact, in the molecular scale, the protein structure of a single ion channel can be analyzed through x-ray crystallography, cryo-electron microscopy (cryo-EM), or high-resolution NMR techniques \cite{Do1998,Mac2003,Ca2010,SMSS2018,GT2020,MZG2021, WYYBR2025}. It indicated that ions are transported through a single ion channel with a tiny pore. The pore size of a single ion channel ranges from $1 $\AA~to $15 $\AA~\cite{AH1998,NMPD2016,KRST2019}, while some special large conductance ion channels can reach the size of about $40$\AA~(4 nm)~\cite{CMDM1997}. The nano- and subnano-scale ion channels, with their atomic-scale selectivity filters, restrict ions to flow through channels individually \cite{Do1998,BR2001,MZM2001,ZMCKM2001}. Under such a subnano-scale, quantum mechanical effects, such as quantum coherence or tunneling effect, may become relevant and essential in the neural signal transmission \cite{VP2010, HHCSC2010, SSB2012, SNS2017}. Furthermore, the free energy profiles along the permeation pathway have been calculated from molecular dynamics simulations. From the analysis of the free energy landscape, the ions perform single-file permeation, and they are confined within binding sites separated by energy barriers \cite{BR2001}. The high sensitivity of ion motion to the details of potential energy suggests that continuous classical models may be insufficient to capture the full ionic dynamics. This motivates a quantum mechanical treatment of ion transport in ion channels. In the literature, the dynamics of ion transmission have been investigated utilizing the classical Brownian motion and the diffusion equation \cite{RABI,MWK,MBYWA2012}. However, a general quantum framework for ion channel dynamics remains to be developed. Moreover, the ion channel current measured in experiments ranges from picoamperes to nanoamperes. This scale corresponds to electronic currents in the low-dimensional nanoelectronic devices, such as two-dimensional (2D) materials or quantum point contacts \cite{HJ1996}. Crucially, both systems involve single-particle transport processes: the ions traverse in ion channels and the electrons transport in nanoelectronic devices. This similarity suggests that quantum transport frameworks of nano-electronic systems may provide insights into ion channel dynamics. Thus, we propose that the quantum transport process may happen in the ion channel.

\begin{figure*}
	\centering
	\includegraphics[width=0.8\linewidth]{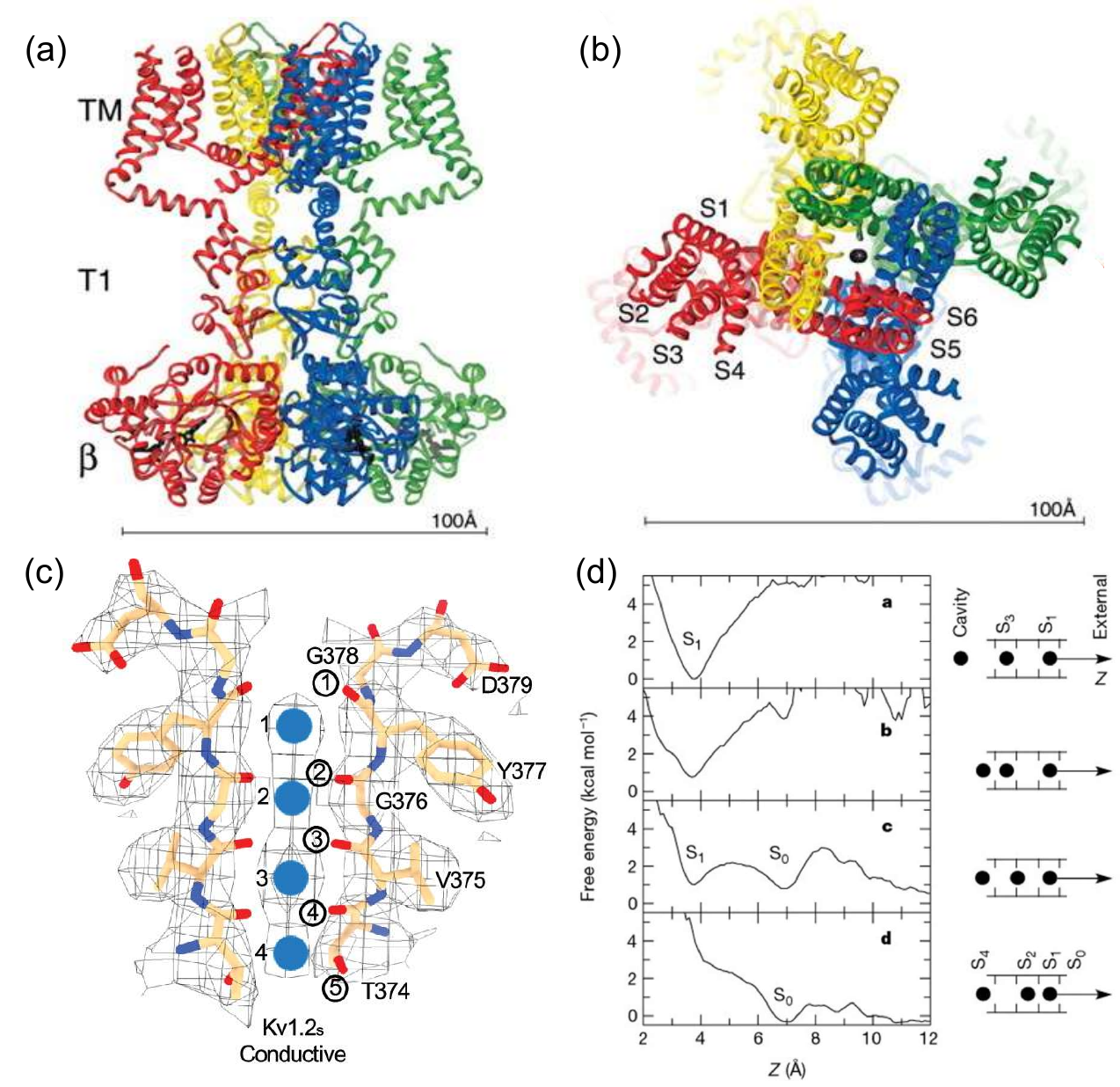}
	\caption{ (color online) Schematic of the Kv1.2 channel architecture and the ion confinement in the selectivity filter.
		(a) The protein structure of a Kv1.2 channel \cite{LCM}. (Reproduced from Fig.~2(A) of S. B. Long et al., Crystal Structure of a Mammalian Voltage-Dependent Shaker Family $K^+$ Channel, Science, © 2005 by The American Association for the Advancement of Science)
		(b) The top view of the transmembrane region. The protein helix of S1 to S4 is the voltage-sensing region, and the pore enclosed by S5 and S6 is the selectivity filter(SF) \cite{LCM}. (Reproduced from Fig.~2(C) of Stephen B. Long et al., Crystal Structure of a Mammalian Voltage-Dependent Shaker Family $K^+$ Channel, Science, © 2005 by The American Association for the Advancement of Science)
		(c) The amino acid structure and the binding sites of the SF region. It is the narrowest region, which consists of four repeating sequences of TVGYG amino acids, allowing potassium ions to stably occupy \cite{WYYBR2025}. (Reproduced from Fig.~4(A) of Y. Wu et al., CryoEM structures of Kv1.2 potassium channels, conducting and non-conducting, (2025) eLife, with the permission from author.)
		(d) Free energy along the selectivity filter obtained from the molecular dynamics simulation, illustrating discrete binding sites and the confinement of ions in the energy well \cite{BR2001}. (Reproduced from Fig.~4 of S. Bernèche et al., Energetics of ion conduction through the $K^+$ channel, Nature, © 2001 by Springer Nature)} 
	\label{fig1:ion channel}
\end{figure*}

The narrowest region of the ion channel is the selectivity filter. Ions inside the selectivity filter are in single-file arrangements due to its specific structure. As a concrete example, consider a voltage-gated potassium channel Kv1.2 \cite{LCM}; see Fig.~\ref{fig1:ion channel}(a). It is composed of the tetramer protein structure. Each subunit contains a voltage-sensing domain formed by the S1 through S4 helices and contributes to the selectivity filter region formed by the S5 and S6 segments; see Fig.~\ref{fig1:ion channel}(b). The selectivity filter is formed by the amino acid sequence TVGYG \cite{KPH, WYYBR2025}; see Fig.~\ref{fig1:ion channel}(c). Molecular dynamics simulations and free energy calculations reveal that the sites inside the selectivity filter are free energy minima, which are called "binding sites". The energy barriers of ions transporting onto the binding sites are approximately $2-4$ kcal/mol ($\sim 90-170$ meV) \cite{BR2001}; see Fig.~\ref{fig1:ion channel}(d), which is significantly larger than the thermal energy $k_B T\sim 25$ meV at physiological temperature. This energy barrier height indicates that the thermal activation over barriers is suppressed, and suggests that quantum mechanical effects, including coherent tunneling between sites, may dominate the transport. At these sites, ions experience minimal interaction energy with neighboring ions and amino acid residues, enabling stable localization. The discrete occupancy of binding sites, with each site accommodating at most only a single ion, bears strong conceptual similarity to Coulomb blockade in nanoelectronic devices. This motivation supports a quantum transport description within the selectivity filter. The binding sites correspond to local minima in the potential energy landscape, which can be treated as discrete ionic energy states. Consequently, a general quantum ion channel model should be a multi-level tunneling quantum system, with the intracellular and extracellular ionic solutions serving as reservoir leads coupled by an external bias voltage. A similar model corresponding to the Coulomb blockade phenomenon in the selectivity filter has been discussed in previous theoretical work \cite{KME}. The discrete binding sites in the ion channel mediate ion transport between ionic solutions (reservoirs). It can be analogous to a molecular junction where discrete molecular orbitals mediate electron transport between metallic electrodes. Thus, a single ion channel can be treated as a quantum molecular junction conceptually. 

To implement this idea, we propose a multi-level quantum tunneling junction model to depict such a molecular junction. The quantum electronic tunneling systems in nanoelectronics typically consist of a double-barrier structure coupled with two electrode leads under an applied external voltage \cite{HJ1996}. The central region between the double barriers is described as an artificial atom with quantized energy levels. In the last two decades, we have developed a systematic time-dependent non-equilibrium quantum transport theory \cite{TuZhang2008,JTZY, LZ2012,YLZ,YZ2017,YLZ2023} for nanoelectronics and quantum materials. This allows us to extend this framework to analyze the ion transport phenomenon in the nervous system microscopically. Within this dynamical transport theory, the ion tunnelings through the ion channel and the associated memory effects can be analyzed quantum mechanically. In particular, the memory effects in the tunneling dynamics can be captured by the dissipative memory kernel and the noise (fluctuations with memory) in the non-Markovian quantum Langevin equation \cite{YLZ}, which characterizes the general non-Markovian behavior of open quantum systems and the non-Markovian nature arises from retarded feedback through the reservoirs \cite{ZLXTN,WMZ2019}. Here, the intracellular and extracellular ionic solutions near the ion channel serve as the two-terminal reservoir leads, and the memory kernel of the non-Markovian quantum Langevin equation is determined by the strength and the time scale of ions in the intracellular and extracellular ion solutions coupled to the ions in the ion channel. This framework enables us to investigate the quantum memory of ion dynamics in the ion channel from the very fundamental quantum level.

To explore neural signal transmissions in ion channels, we utilize our time-dependent non-equilibrium quantum transport theory 
\cite{JTZY, LZ2012,YLZ,YZ2017} to study the dynamics of the ion channel current and the differential conductance. To mimic the stimulation of neurons from the outside, we apply external time-varying voltages across the ionic junction as the time-dependent cross-membrane potentials. In particular, we extend the time-dependent transport theory of mesoscopic electronic systems \cite{JTZY} to ion channels, where the external voltages drive photon-assisted resonant tunneling processes. Within this framework, we find that the time-dependent voltage modulates the tunneling amplitudes, introducing a dynamical phase formally equivalent to an effective magnetic flux phase. This phase modulation sustains quantum coherence in the ion tunneling dynamics and introduces time-dependent memory effects in ion transport. We term this memory effect the active quantum memory because it is derived by the external time-varying voltage. Active quantum memory manifests in the ion transport as a persistent oscillating current in the ion channel, referred to as the sideband effect \cite{WJM,JWM,JTZY}. The sideband current serves as a signature of the system carrying quantum memory and quantum coherence, and it forms a hysteresis loop with multiple crossing points in the I-V curve. We will analyze the dependence of the sideband current on the time-varying voltage between the intracellular and extracellular ionic solutions, the energy levels of ion sites in the ion channel, and the chemical potentials of the intracellular and extracellular ionic solutions. We find that the tunneling effect is enhanced when the chemical potentials match the renormalized energy levels. Under the strongest resonance, the multi-level system exhibits transport characteristics of an effective single-level resonant tunneling junction. The crossing points in the current-voltage hysteresis (I-V curve) and the differential conductance (G-V diagram) indicate the coherence tunneling and may serve as a measure of the active quantum memory effect in the nervous system in general. Moreover, the temperature dependency of the active quantum memory and its application to the real biological scale are also explored in this work. For the practical applications, such a theory for quantum memory circuits may be valuable for the development of neuromorphic computing technology in the future.

The rest of the paper is organized as follows. In Sec.~II, motivated by the structure of ion channels, we introduce the multi-level ionic tunneling junction. Ion tunneling through the junction carries time-dependent quantum phases associated with an effective magnetic flux induced by the time-varying voltage. As a concrete example, we introduce a four-level ionic tunneling model to describe the selectivity filter of a Kv1.2 ion channel and derive the corresponding transport current from the first principle of quantum mechanics rigorously. In Sec.~III, we present analytic and numerical results assuming a wide-band spectral density for the intracellular and extracellular ionic solutions. After applying time-varying voltages across the ion channel, we formulate the corresponding transient current in detail. The ion channel current displays sideband oscillations under the time-varying voltage, which produce a multi-crossing hysteresis loop in the I-V characteristics. These features serve as signatures of quantum coherence and active quantum memory induced by the time-varying voltages. We identify the effective magnetic flux strength as a critical parameter for quantifying the degree of active quantum memory, and the number of crossing points can serve as a measurement of the amount of memory. The steady-state differential conductance is also analyzed under varying degrees of active quantum memory. We also examine how temperature affects the quantum memory effect and why the thermal fluctuations are suppressed in the quantum coherent transport of ion channel dynamics. Finally, we apply our framework to a realistic ion channel system with experimentally relevant parameters. The discussions and perspectives are presented in Sec.~IV.

\section{Modeling ion channel as multi-level quantum junction}
	
\subsection{A quantum modeling for ion channel}
Ion channels are embedded in the membrane, which connects the intracellular and extracellular ionic solutions. The protein structures of many voltage-gated ion channels are now very clear, especially those in the Kv channel family. Figure~\ref{fig1:ion channel}(a) demonstrates the protein structure of a voltage-gated Kv1.2 potassium ion channel \cite{LCM}. The Kv channel family shares a common structural motif, with each member comprising four subunits assembled into a tetramer. The structure of the whole ion channel contains the transmembrane pore, the T1 domain, and the $\beta$ subunit. Here, we want to focus on the segment of the transmembrane region, the pore formed by the protein helices S5 and S6, which is also the location of the selectivity filter \cite{LCM}. The relative position of the selectivity filter, surrounded by S5 and S6, is clearly visible in the top view of the transmembrane region shown in Fig.~\ref{fig1:ion channel}(b). Figure~\ref{fig1:ion channel}(c) shows the diagrams of electron density distributions from the molecular simulation 
dynamics \cite{WYYBR2025}, from which we can understand how ions arrange in a single file in the selectivity filter. The sites where ions are bound in the selectivity filter are low-energy binding sites that allow ions to remain bound stably \cite{WYYBR2025}. In molecular dynamics simulations, Bernèche and Roux simulated ions moving through the selectivity filter \cite{BR2001}. One can see that the ion is confined in the binding site by the free energy well, and the energy map transforms as ions move; see Fig.~\ref{fig1:ion channel}(d). The energy wells are on the order of a few kcal $\rm{mol}^{-1}$ (tens to hundreds of meV).

From our perspective, the binding sites within the selectivity filter are analogous to a molecular or artificial atom with quantized energy levels. To construct a quantum physical model for ion channels, we treat each ion channel as a molecular or atomic tunneling junction. By connecting two "ionic reservoir leads" (corresponding to the intracellular or extracellular ionic solutions) through the coupling with a nanoscale ion channel, we model the ion channel as an ionic tunneling junction with quantized energy levels. The general Hamiltonian describing the quantum transmission of ions through the ion channels can have the form,
\begin{align}
	\hat{H}(t) = \sum_{ij} & \epsilon_{ij} \hat{a}_{i}^\dag \hat{a}_{j} + \sum_{\alpha k} \epsilon_{\alpha k}(t)\hat{c}_{\alpha k}^\dag \hat{c}_{\alpha k} \notag\\ 
	&+ \sum_{i\alpha k} (V_{i\alpha k} \hat{a}_{i}^\dag \hat{c}_{\alpha k} + V^*_{i\alpha k}\hat{c}_{\alpha k}^\dag \hat{a}_{i}),
	\label{eq:Hamiltonian}
\end{align}
where the first term represents the energy levels of the sites inside the ion channel and the transition between sites. The ion channel can contain a few (say $N$) discrete energy levels $\epsilon_{ii}$ (with $i =1,2,...,N$), and the intersite transition amplitude $\epsilon_{ij,\{i\neq j\}}$. The operators $\hat{a}_{i}^\dag (\hat{a}_{i})$ are the creation (annihilation) operators of each energy state in the ion channel.  The second term describes the ion Hamiltonian for the intracellular and extracellular ionic solutions denoted by $\alpha = I, E$ respectively. The operators $\hat{c}^\dag_{\alpha k}(\hat{c}_{\alpha k})$ are the creation (annihilation) operators of the $k^{th}$ energy state of ionic charge carriers in the ionic solution $\alpha$, and $\epsilon_{\alpha k}$ is the corresponding quantum energy level (usually all the energy levels form a continuous spectrum). By defining $\epsilon_{\alpha k}(t) = \epsilon_{\alpha k} + qV_\alpha(t)$, the ionic solution levels can be adjusted through the external time-varying voltage (potential) $V_\alpha(t)$, where $q$ is the charge of each ion. Note that in the neuronic ion channels, the transport ions are typically the sodium and potassium ions, $Na^+$ and $K^+$, both carry a positive charge $q$. The bias voltage, $V_b(t)=V_I(t)-V_E(t)$ represents the time-varying  potential difference between the intracellular or extracellular ionic solutions near the membrane, caused by the stimulations of neurons from the outside. The third term denotes the tunneling interaction of ions between the central sites and the ionic solutions, and $V_{i\alpha k}$ is the corresponding tunneling amplitude. Also note that $Na^+$ and $K^+$ have a half-integer spin ($s=3/2$) so that they can be treated as fermionic particles. Thus, all the creation and annihilation operators, $\hat{a}_{i}^\dag (\hat{a}_{i})$ and $\hat{c}^\dag_{\alpha k}(\hat{c}_{\alpha k})$, obey the anti-commutation relations and the ion occupations obey the Fermi-Dirac statistical distribution in the equilibrium state.

\begin{figure}
	\centering
	\includegraphics[width=\linewidth]{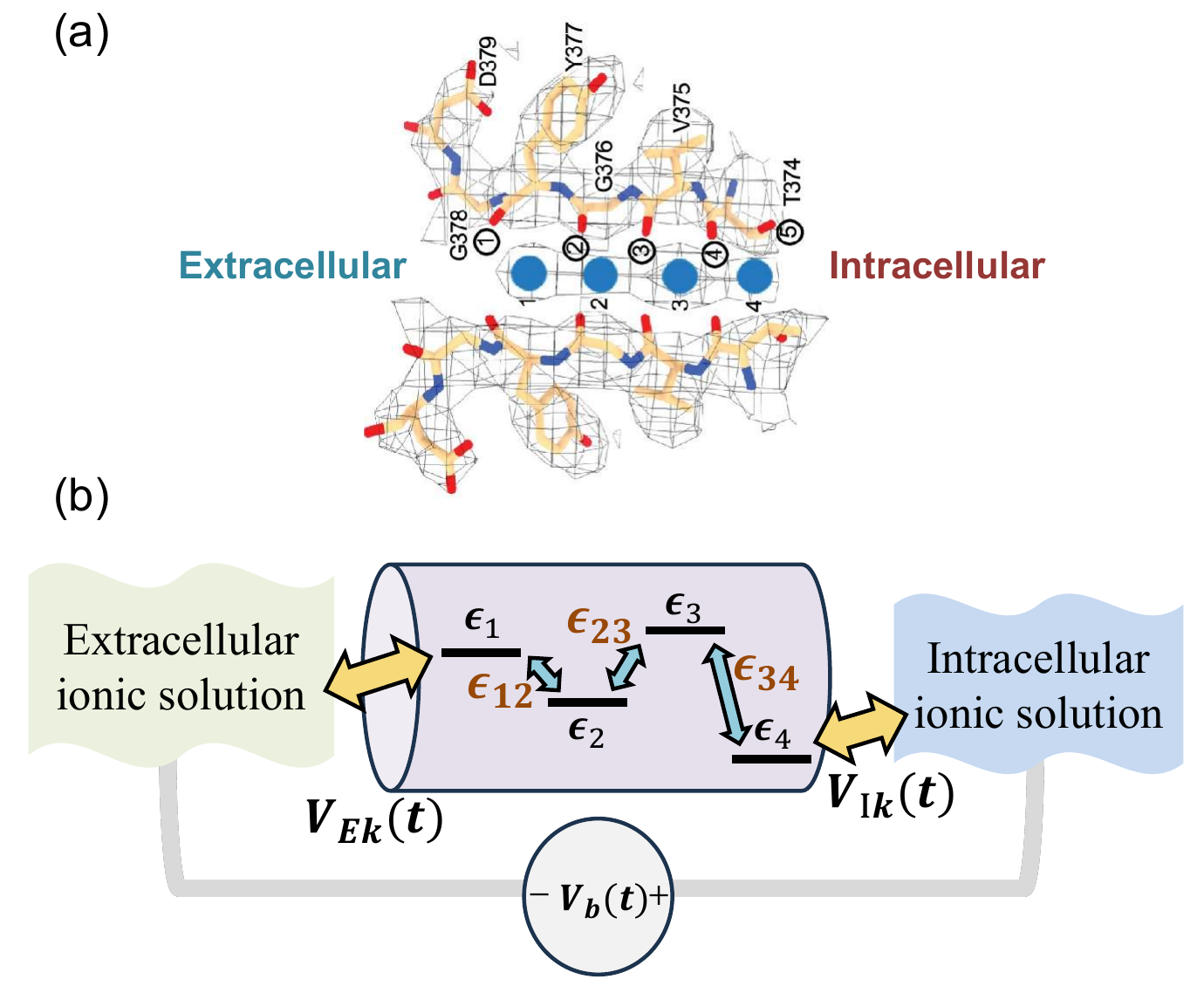}
	\caption{(color online) 
			Schematic plots for the connection of the selectivity filter of the Kv1.2 channel with the quantized multi-level tunneling junction. (a) The lateral direction of the selectivity filter is chosen for the Kv1.2 channel \cite{WYYBR2025}. (Modified from Fig.~4(A) of Y. Wu et al., CryoEM structures of Kv1.2 potassium channels, conducting and non-conducting, (2025) eLife, with the permission from author.) (b) The ionic tunneling system with discrete quantum levels is modeled as the quantum structure of the selectivity filter. The four-level ionic tunneling junction connects with the extracellular and intracellular ion solutions. The discrete quantum levels model the binding sites, and the time-varying membrane potential difference is taken as the time-varying bias voltage $V_b(t)=V_I(t)-V_E(t)$.} 
	\label{fig2:Kv1.2 ion channel}
\end{figure}

To construct a specific quantum ion channel model, we analogize the structure of the Kv1.2 channel. As shown in Fig.~\ref{fig2:Kv1.2 ion channel}(a), the Kv1.2 channel is turned to the lateral direction, and the four binding sites in the ion channel correspond to four independent energy levels $\{\epsilon_{i}\}$, $i=1,...,4$ in the molecular junction. A schematic diagram for the quantized ion channel model is shown in Fig.~\ref{fig2:Kv1.2 ion channel}(b). Thus, the Hamiltonian of Kv1.2 channel is reduced to the following form,
\begin{alignat}{2}
	\hat{H}(t) =&\sum_{i=1}^4 \epsilon_i \hat{a}_i^\dag \hat{a}_i + (&&\epsilon_{12}^* \hat{a}_2^\dag \hat{a}_1 + \epsilon_{23}^*\hat{a}_3^\dag \hat{a}_2 + \epsilon_{34}^* \hat{a}_4^\dag \hat{a}_3 \notag\\
	&  &&+\epsilon_{12} \hat{a}_1^\dag \hat{a}_2 + \epsilon_{23}\hat{a}_2^\dag \hat{a}_3 + \epsilon_{34} \hat{a}_3^\dag \hat{a}_4) \notag\\
	&+ \sum_{\alpha k} \epsilon_{\alpha k}(t)&&\!\!\!\hat{c}_{\alpha k}^\dag \hat{c}_{\alpha k} \notag\\ 
	&+\sum_k(V_{1Ek} \hat{a}_1^\dag &&\!\hat{c}_{Ek}+ V^*_{1Ek}\hat{c}_{Ek}^\dag \hat{a}_1\notag\\
	&  &&+ V_{4Ik}\hat{a}_4^\dag \hat{c}_{Ik} + V^*_{4Ik}\hat{c}_{Ik}^\dag \hat{a}_4).
	\label{eq:Kv1.2 Hamiltonian}
\end{alignat}
Here, we consider a nearest-neighbor coupling where intersite transitions occur only between adjacent sites. Additionally, the extracellular ionic leads are set to couple to the first site ($i=1$) and the intracellular ionic lead couples to the last site ($i=4$). Therefore, only two couplings, $V_{1Ek}$ and $V_{4Ik}$, dominate the tunneling to the ion solutions, while the remaining couplings $V_{i\alpha k}$ in the general Hamiltonian of Eq.~(\ref{eq:Hamiltonian}) vanish. As a result, ion transport is induced by the couplings of the extracellular ion solution to the first site and the intracellular ion to the last site through the intersite transitions in the ion channel in the lateral direction, as shown schematically in Fig.~\ref{fig2:Kv1.2 ion channel}(b).

\subsection{Time-dependent nonequilibrium quantum transport theory for the memory dynamics in ion channel}
Now, we can study the ion channel dynamics from the first principle of quantum mechanics. 
Ion transport is driven by the time-varying voltage $V_b(t)$ across the membrane, which changes the energy landscape 
of the binding sites. The time-dependent ionic current through the ion channel is expressed by
\begin{align}
	I(t) = I_E(t) - I_I(t), 
\end{align}
where $I_I(t)$ ($I_E(t)$) is the current of ions flowing from the ion channel to the intracellular (extracellular) ionic solution. It is defined as the time derivative of the total ion charges in the intracellular (extracellular) ionic solution. 
Explicitly, using the Heisenberg equation of motion in quantum mechanics, we have
\begin{align}
	I_\alpha(t)& \equiv q\Big\langle \frac{d}{dt} \hat{N}_\alpha(t)\Big\rangle = \frac{q}{i\hbar}\langle [\hat{N}_\alpha, \hat{H}]\rangle \notag \\
	&= \frac{iq}{\hbar}\sum_{ik}\Big[V_{i\alpha k}\langle \hat{a}_i^\dagger(t)\hat{c}_{\alpha k}(t)\rangle - V^*_{i\alpha k}\langle \hat{c}^\dagger_{\alpha k}(t) \hat{a}_i(t)\rangle\Big],
	\label{eq:current}
\end{align}
where $\alpha = I, E$ labels the intracellular and extracellular ionic solutions, respectively, and $q$ is the charge carried by each ion. 
The notation $\langle \hat{O}(t) \rangle = {\rm{Tr}}[\hat{O}(t)\rho_{tot}]$  denotes the expectation value of any quantum observable $\hat{O}(t)$ in the Heisenberg picture of quantum mechanics. 

In the Heisenberg picture, quantum states do not change in time, namely $\rho_{\rm tot}(t)=\rho_{\rm tot}(t_0)\equiv\rho_{\rm tot}$. 
The time-evolution of the system is determined by the Heisenberg equation of motion, $\frac{d}{dt}\hat{O}(t)=\frac{1}{i\hbar}[\hat{O}(t),\hat{H}(t)]$.
Without loss of generality, we can take the initial state of the whole system (the ion channel plus the ionic solutions surrounding the membrane) as given by the density matrix 
\begin{align}
\rho_{tot} = \rho_{tot}(t_0) = \rho_s(t_0) \otimes \rho_I \otimes \rho_E,
\end{align} 
where the ion channel can be initially in any arbitrary quantum state $ \rho_s(t_0)$, and the intracellular and extracellular ionic solutions are initially in the thermal equilibrium state, 
\begin{align}
\rho_\alpha=\frac{1}{Z_\alpha}e^{-\frac{1}{k_BT} \sum_{k} \epsilon_{\alpha k}(t_0) \hat{c}_{\alpha k}^\dag \hat{c}_{\alpha k} }
\end{align} 
in which $k_B$ is the Boltzmann constant, $T$ is the thermal temperature of ionic solution, and $Z_\alpha$ is the corresponding thermal partition function in statistical mechanics. 

Using the Heisenberg equation of motion, we obtain the equations of motion for operators $\hat{a}_{i}(t)$ and $\hat{c}_{\alpha k}(t)$ in Eq.~(\ref{eq:current}) as follows,
\begin{subequations}
	\begin{align}
		& \frac{d\hat{a}_{i}(t)}{dt}  = -\frac{i}{\hbar}\sum_j\epsilon_{ij} \hat{a}_{j}(t) - \frac{i}{\hbar}\sum_{\alpha k}V_{i\alpha k}\hat{c}_{\alpha k}(t),
		\label{eq:dif_a}  \\
		& \frac{d\hat{c}_{\alpha k}(t)}{dt}  = -\frac{i}{\hbar}[\epsilon_{\alpha k} + qV_\alpha(t)] \hat{c}_{\alpha k}(t) - \frac{i}{\hbar}\sum_{j}V^*_{j\alpha k}\hat{a}_{j}(t).
		\label{eq:dif_b}
	\end{align}
\end{subequations}
The operator  $\hat{c}_{\alpha k}$ in Eq.~(\ref{eq:dif_b}) 
can be formally solved,
\begin{align}
	\begin{split}
		\hat{c}_{\alpha k}(t) =& e^{-\frac{i}{\hbar}\int_{t_0}^tdt'[\epsilon_{\alpha k} + qV_\alpha(t')]}\hat{c}_{\alpha k}(t_0)\\
		&- \frac{i}{\hbar}\sum_{j}\int_{t_0}^t dt' V_{j\alpha k}^* e^{-\frac{i}{\hbar}\int_{t'}^t d\tau(\epsilon_{\alpha k} + qV_\alpha(\tau))}\hat{a}_{j}(t').
	\end{split}
	\label{ionsolutions}
\end{align}
Substituting the above solution into Eq.~(\ref{eq:dif_a}), we obtain the following exact quantum Langevin equation in 
the on-site ion state operators $\hat{a}_{i}(t)$ \cite{YLZ}, 
\begin{equation}
	\frac{d\hat{a}_i(t)}{dt} + \frac{i}{\hbar}\sum_j \epsilon_{ij}\hat{a}_j(t) + \int_{t_0}^t dt' \sum_j g_{ij}(t,t')\hat{a}_j(t') =  \hat{f}_i(t),
	\label{eq:QLE1}
\end{equation}
where $\epsilon_{ij}$ are the on-site energy levels and the transition energies between different levels in the ion channel, 
and $g_{ij}(t,t')$ and 
$\hat{f}_i(t)$ are respectively the dissipative memory kernel and the noise force in the quantum Langevin equation 
that are derived exactly from the first principle of quantum mechanics,
\begin{subequations}
\label{mknf}
\begin{align}
	g_{ij}(t,t')  = \frac{1}{\hbar^2} \sum_{\alpha k} V_{i\alpha k}V^*_{j\alpha k} e^{-\frac{i}{\hbar}\int_{t'}^td\tau[\epsilon_{\alpha k} + qV_\alpha(\tau)]},  \label{mk} \\
	\hat{f}_i(t) = -\frac{i}{\hbar}\sum_{\alpha k}V_{i\alpha k}e^{-\frac{i}{\hbar}\int_{t_0}^tdt'[\epsilon_{\alpha k} + qV_\alpha(t')]}\hat{c}_{\alpha k}(t_0). 	\label{nf}	
\end{align}
\end{subequations}
The above quantum Langevin equation of Eq.~\eqref{eq:QLE1} is a time-convolution equation of motion, which describes the quantum dissipative and fluctuational dynamics of ions in the ion channel. 
Thus, solving the above exact quantum Langevin equation, we can obtain the dynamics of ion transport current given by 
Eq.~(\ref{eq:current}) and see how the diffusion and stochastic dynamics of the ion motion with memory 
emerge in the ion channel. 

For convenience,  the exact quantum Langevin equation Eq.~(\ref{eq:QLE1}) may be rewritten in a matrix form, 
\begin{equation}
	\frac{d}{dt}\hat{{\boldsymbol{a}}}(t) + \frac{i}{\hbar}\boldsymbol{\epsilon}_s\hat{\boldsymbol{a}}(t) + \int_{t_0}^tdt'\boldsymbol{g}(t,t')\hat{\boldsymbol{a}}(t') =  \hat{\boldsymbol{f}}(t). 	\label{eq:QLE}
\end{equation}
For the Kv1.2 channel model, the operator vector $\hat{\boldsymbol{a}}(t)=\{\hat{a}_1(t), \hat{a}_2(t), \hat{a}_3(t), \hat{a}_4(t)\}$ represents the four on-site energy levels in the ion channel, and $\boldsymbol{\epsilon}_s$ is a $4\times 4$ matrix with matrix element $\epsilon_{ij}$ for the energy level and the transition energy between different sites, respectively. From Eq.~(\ref{eq:Kv1.2 Hamiltonian}), this energy matrix is given explicitly by
\begin{align}
	\bm{\epsilon_s} &= 
	\begin{bmatrix}
		\epsilon_1 & \epsilon_{12} & 0 & 0 \\
		\epsilon_{12}^* & \epsilon_2 & \epsilon_{23} & 0 \\
		0 & \epsilon_{23}^* & \epsilon_3 & \epsilon_{34} \\
		0 & 0 & \epsilon_{34}^* & \epsilon_4
	\end{bmatrix}.
	\label{bs}
\end{align}
The dissipative memory kernel $\boldsymbol{g}(t,t')$ in Eq.~(\ref{eq:QLE}) is also a $4\times 4$ matrix with elements
\begin{align}
	\bm{g}(t,t') &= 
	\begin{bmatrix}
		g_E(t,t') & 0 & 0 & 0\\
		0 & 0 & 0 & 0 \\
		0 & 0 & 0 & 0 \\
		0 & 0 & 0 & g_I(t,t')		
	\end{bmatrix} .  \label{gtt}
\end{align}
This simple dissipative memory kernel is given for the Kv1.2 channel model, because only the coupling strengths $V_{1Ek}$ and $V_{4Ik}$ are nonzero, as specified in 
Eq.~\eqref{eq:Kv1.2 Hamiltonian}. Hence, $\boldsymbol{g}(t,t')$ only contains two nonzero elements,
\begin{subequations}
\label{mk1}
\begin{align}
	&g_E(t,t') = \frac{1}{\hbar^2} \sum_{k} V_{1E k}V^*_{1 E k} e^{-\frac{i}{\hbar}\int_{t'}^td\tau[\epsilon_{E k} + qV_E(\tau)]},  \\
	&g_I(t,t') = \frac{1}{\hbar^2} \sum_{k} V_{4I k}V^*_{4I k} e^{-\frac{i}{\hbar}\int_{t'}^td\tau[\epsilon_{I k} + qV_I(\tau)]}.  
\end{align}
\end{subequations}
These two dissipative memory kernels in the quantum Langevin equation represent the non-Markovian memories describing historical ionic back reactions between the central ion channel region and the extracellular and intracellular ionic solutions, respectively. The ionic back reactions characterize the detailed path of ion transport, i.e.~ions tunnel from the ion channel into the ionic solutions at time $t$ and then tunnel back into ion channel at time $t'$.
The vector expression of the noise force in Eq.~(\ref{eq:QLE}) is also simply given by
\begin{align}
	\hat{\bm{f}}(t) = 
	\begin{bmatrix}
		\hat{f}_E(t)\\ 0\\ 0\\ \hat{f}_I(t)
	\end{bmatrix} .
\end{align}
They are arisen from the intracellular and extracellular ionic solutions, respectively, 
\begin{subequations}
\begin{align}
	& \hat{f}_E(t) = -\frac{i}{\hbar}\sum_{ k}V_{1E k}e^{-\frac{i}{\hbar}\int_{t_0}^tdt'[\epsilon_{E k} + qV_E(t')]}\hat{c}_{E k}(t_0), \label{nf1E}\\
	&\hat{f}_I(t) = -\frac{i}{\hbar}\sum_{ k}V_{4I k}e^{-\frac{i}{\hbar}\int_{t_0}^tdt'[\epsilon_{I k} + qV_I(t')]}\hat{c}_{I k}(t_0). 
	\label{nf1I}
\end{align}
\end{subequations}
The origin of noise forces of Eqs.~(\ref{nf1E}) and (\ref{nf1I}) comes from the thermal ions which initially live in the thermal-equilibrium
intracellular and extracellular ionic solutions.

Because of the linearity of the equation, the general solution of Eq.~(\ref{eq:QLE}) can be expressed as \cite{YLZ}
\begin{align}
	\hat{\boldsymbol{a}}(t) = \boldsymbol{u}(t,t_0)\hat{\boldsymbol{a}}(t_0) + \hat{\boldsymbol{F}}(t),    \label{ion_channel}
\end{align}
where $\boldsymbol{u}(t,t_0) \equiv \langle \{ \hat{\bm a}(t), \hat{\bm a}^\dag(t_0) \} \rangle $ is called the generalized  retarded Green function matrix in the non-equilibrium dynamics, and $\hat{\boldsymbol{F}}(t)$ is a noise field induced by the noise force \cite{JTZY,YLZ,YZ2017}. Both of them determine the ion dynamics in the ion channel through ion back reactions between the central region of the ion channel and the intracellular and extracellular ionic solutions. 
Substituting this solution into Eq.~(\ref{eq:QLE}), we can show that the non-equilibrium Green function $\boldsymbol{u}(t,t')$ and the noise field $\hat{\boldsymbol{F}}(t)$ obey the following homogeneous and inhomogeneous time-convolution equations of motion,
\begin{subequations}
	\label{convoeq}
	\begin{align}
		& \frac{d}{dt}\boldsymbol{u}(t,t_0) + \frac{i}{\hbar}\boldsymbol{\epsilon}_s\boldsymbol{u}(t,t_0) + \int_{t_0}^t dt'\boldsymbol{g}(t,t')\boldsymbol{u}(t',t_0) =0,
		\label{eq:u} \\
		& \frac{d}{dt}\hat{\boldsymbol{F}}(t) +\frac{i}{\hbar}\boldsymbol{\epsilon}_s\hat{\boldsymbol{F}}(t) + \int_{t_0}^t dt' \boldsymbol{g}(t,t')\hat{\boldsymbol{F}}(t') = \hat{\boldsymbol{f}}(t),
		\label{eq:F}
	\end{align}
\end{subequations}
subjected to the boundary condition $\boldsymbol{u}(t_0,t_0)=\mathbb{I}_4$ and the initial condition $\hat{\boldsymbol{F}}(t_0)=0$. 
As a result of Eq.~(\ref{eq:QLE}), the integro-differential equations of Eqs.~(\ref{eq:u}) and (\ref{eq:F}) are also the time-convolution equations of motion that encompass all possible dissipative and fluctuational memory dynamics arisen from the ion back reactions and are manifested through the time-convolution integrals.  

Furthermore, the noise field of Eq.~(\ref{eq:F}) has the analytical solution,
\begin{align}
	\hat{\boldsymbol{F}}(t) = \sum_{\alpha}\int_{t_0}^t dt'\boldsymbol{u}(t,t')\hat{\boldsymbol{f}}_\alpha(t').  \label{noisef}
\end{align}  
It is easy to show that the noise field induces the following noise correlation functions, 
\begin{subequations}
	\label{noises}
	\begin{align}
		  \langle \hat{\bm F} (t) \rangle  & \equiv \Tr[\hat{\bm F}(t) \rho_{\rm tot}]=0 , \label{eq:aveF}\\
		  \langle \hat{\bm F}^\dag_j(t) & \hat{\bm F}_i(t') \rangle  
		= \! \int_{t_0}^{t'} \!\! d\tau \!\! \int_{t_0}^t \!\!\! d\tau' [{\bm u}(t',\tau)\widetilde{\bm g}(\tau,\tau'){\bm u}^\dag(t,\tau')]_{ij} \notag \\
		&=  \boldsymbol{v}_{ij}(t',t) ,
		\label{eq:FDT}
	\end{align}
\end{subequations}
where $\widetilde{\bm g}(t, t')$ is the fluctuational (noise-correlation) memory kernel defined by,
\begin{align}
	[\widetilde{\bm g}(t,& t')]_{ij}  = \notag \\
	& \frac{1}{\hbar^2} \sum_{\alpha k} V_{i\alpha k}V^*_{j\alpha k} f_\alpha(\epsilon_{\alpha k}) e^{-\frac{i}{\hbar}\int_{t'}^td\tau[\epsilon_{\alpha k} + qV_\alpha(\tau)]},  \label{ckm} 
\end{align}
and $f_\alpha(\epsilon_{\alpha k})= 1/[1+e^{\frac{1}{k_B T}(\epsilon_{\alpha_k} - \mu_\alpha)}]$ is the initial ion number distribution 
(the Fermi-Dirac distribution) of the ionic solution $\alpha$ with $\mu_\alpha$ being the corresponding ionic chemical potential. 
The fluctuational memory kernel matrix of the noise correlation functions $\widetilde{\bm{g}}(t,t')$ is given by
\begin{align}
	\widetilde{\bm{g}}(t,t') &= 
	\begin{bmatrix}
		\widetilde{g}_E(t,t') & 0 & 0 & 0\\
		0 & 0 & 0 & 0 \\
		0 & 0 & 0 & 0 \\
		0 & 0 & 0 & \widetilde{g}_I(t,t')		
	\end{bmatrix}.
	\label{eq:gt_Matrix}
\end{align}
In fact, Eqs.~(\ref{eq:aveF}), (\ref{eq:FDT}) and (\ref{ckm}) fully describe the general noise-induced fluctuational memory dynamics. It also gives a fundamental derivation of the nonequilibrium fluctuation-dissipation theorem for any open quantum system \cite{JTZY, LZ2012,YLZ,YZ2017,YLZ2023,ZLXTN, WMZ2019}.

Substituting the solution of Eqs.~(\ref{ionsolutions}) and (\ref{ion_channel}) into the ionic transport current of Eq.~\eqref{eq:current},  
the intracellular and extracellular ion currents are obtained  
\begin{subequations}
\label{eq:Kv1.2 current}
\begin{align}
	& I_E(t) = 2q~{\rm Re}\int_{t_0}^t \!\! dt' {\big[\bm{g}(t,t')\bm{n}(t',t) - \widetilde{\bm{g}}(t,t')\bm{u}^\dag(t,t')\big]}_{11}, \label{eq:IE} \\
	& I_I(t) = 2q~{\rm Re}\int_{t_0}^t \!\! dt' {\big[\bm{g}(t,t')\bm{n}(t',t) - \widetilde{\bm{g}}(t,t')\bm{u}^\dag(t,t')\big]}_{44}.
	\label{eq:II}
\end{align}
\end{subequations}
Here $\bm{g}(t,t')$ and $\widetilde{\bm{g}}(t,t')$ are the ionic dissipative and fluctuational memory kernels given respectively by Eqs.~(\ref{gtt}) and (\ref{eq:gt_Matrix}), $\boldsymbol{u}(t,t_0)$ is the retarded Green function which describes the ion dissipation dynamics and is determined by Eq.~(\ref{eq:u}). The matrix function $\boldsymbol{n}(t',t)$ in Eqs.~(\ref{eq:IE}) and (\ref{eq:II}) describes the ionic correlation dynamics in the ion channel, which is defined from the following equation,
\begin{align}
	\boldsymbol{n}_{ij}(t',t) & \equiv \langle \hat{\bm{a}}_j^\dag(t) \hat{\bm{a}}_i(t')\rangle   \notag \\
	&= [\bm{u}(t',t_0)\bm{n}(t_0,t_0)\bm{u}^\dag(t,t_0)]_{ij}  + \bm{v}_{ij}(t',t),
	\label{eq:single_particle}
\end{align} 
where $\bm{n}_{ij}(t_0,t_0) = \langle \bm{a}_j^\dag(t_0) \bm{a}_i(t_0) \rangle$ are the initial particle correlation in the ion channel, 
and $\bm{v}_{ij}(t',t) = \langle \hat{\bm F}^\dag_j(t) \hat{\bm F}_i(t') \rangle $ are the nonequilibrium ion correlated Green functions given by Eq.~(\ref{eq:FDT}). It describes the ion noise correlation and represents the nonequilibrium fluctuation-dissipation relation of ions in the ion channel.

\subsection{Membrane potential induced quantum magnetic flux and active quantum memory}

With the above exact quantum-mechanical formulations and solutions, ion transport through ion channels and the corresponding memory dynamics can now be investigated microscopically.
Indeed, Eqs.~(\ref{eq:IE}) and (\ref{eq:II}) demonstrates the ionic current derived from the model Hamiltonian of Kv1.2 channel, 
which explicitly reveals the physical observable of ion transport dynamics. From Eqs.~(\ref{eq:IE}) and (\ref{eq:II}), one can 
see that the dominant contributions to the ion current originate from the ion dissipation and fluctuation processes in the ion channel, induced from couplings with the intracellular and extracellular ion solutions. They are characterized by the ion dissipative and fluctuational memory kernels, $\boldsymbol{g}(t,t')$ and $\widetilde{\bm{g}}(t,t')$, as well as the time-convolution equation of the ion retarded and correlation Green's function $\boldsymbol{u}(t,t')$ and $\boldsymbol{n}(t,t')$ that involve the ion memory kernels $\boldsymbol{g}(t,t')$ and $\widetilde{\bm{g}}(t,t')$. 
Crucially, the time-varying voltage modulates the memory kernels, giving a controllable memory effect associated with the quantum coherence transport in the ion tunneling processes. 
In the following, we will analyze this voltage-induced memory effect and introduce a quantitative definition of 
the active quantum memory.

In fact, we find that the time-varying voltage $V_\alpha(t)$ induces an effective time-varying magnetic flux,  
\begin{align}
	\Phi_\alpha(t) \equiv \int^t_{t_0} V_\alpha(\tau) d\tau , \label{EMF}
\end{align}
that modifies the ion tunneling amplitude $V_{i\alpha k}$ to carry a time-dependent magnetic flux phase:
\begin{align}
	V_{i\alpha k}(t) = V_{i\alpha k}e^{-i\phi_\alpha(t)},   ~~ V^*_{i\alpha k}(t) =[V_{i\alpha k}(t)]^*.
\end{align} 
The magnetic flux phase $\phi_\alpha(t)$ is associated with the effective magnetic flux $\Phi_\alpha(t)$, 
\begin{align}
	\phi_\alpha(t)= 2\pi \frac{ \Phi_\alpha(t)}{\Phi_0},   \label{mfp}
\end{align} 
and $\Phi_0=h/q$ is the magnetic flux quanta. Physically, the external time-varying potential $V_\alpha(t)$ induces an effective magnetic flux $\Phi_\alpha(t)$ paralleling with the membrane axis such that when the ion tunnels into the ion channel, it carries a quantum phase from this effective magnetic flux. This magnetic flux phase for charge particles moving in a magnetic field cannot be gauged away, according to the general principle of gauge symmetry in quantum mechanics \cite{Tannoudji2020}.

As a result, the quantum memory kernels can be re-expressed in terms of the time-dependent ion tunneling amplitude $V_{i\alpha k}(t)$,
\begin{subequations}
	\label{memo-noisem}
	\begin{align}
		& [\boldsymbol{g}(t,t')]_{ij}  =  \frac{1}{\hbar^2} \sum_{\alpha k} V_{i\alpha k}(t)V^*_{j\alpha k}(t') e^{-\frac{i}{\hbar}\epsilon_{\alpha k}(t-t') },  \label{mkm} \\
		& [\widetilde{\bm g}(t, t')]_{ij}  = \frac{1}{\hbar^2} \sum_{\alpha k} V_{i\alpha k}(t)V^*_{j\alpha k}(t') f_\alpha(\epsilon_{\alpha k})e^{-\frac{i}{\hbar}\epsilon_{\alpha k}(t-t') }.
	\end{align}
\end{subequations}
More explicitly, by introducing the spectral density
$J_\alpha(\epsilon) = 2\pi\sum_k|V_{\alpha k}|^2\delta(\epsilon_{\alpha k} - \epsilon )$ of ionic solution $\alpha$ 
surrounding the membrane, the non-zero elements of the memory kernel matrices $\bm{g}(t,t')$ and $\widetilde{\bm{g}}(t,t')$ given in 
Eqs.~(\ref{gtt}) and (\ref{eq:gt_Matrix}) can be reduced to 
\begin{subequations}
	\label{eq:ggt}
	\begin{align}
		g_\alpha(t,t') & = e^{-i[\phi_\alpha(t)-\phi_\alpha(t')]}   \frac{1}{\hbar^2}  \!\! \int_{-\infty}^{\infty} \!\! \frac{d\epsilon}{2\pi} J_\alpha(\epsilon) e^{-\frac{i}{\hbar}\epsilon(t-t') } \notag \\
		& = M_\alpha (t,t') g^0_\alpha(t,t') , \label{eq:g}  \\
		\widetilde{g}_\alpha(t,t') & = e^{-i[\phi_\alpha(t)-\phi_\alpha(t')]} \frac{1}{\hbar^2}  \!\! \int_{-\infty}^{\infty} \!\! \frac{d\epsilon}{2\pi}J_\alpha(\epsilon)f_\alpha(\epsilon)e^{-\frac{i}{\hbar}\epsilon(t-t') } \notag \\
		& = M_\alpha(t,t') \widetilde{g}^0_\alpha(t,t'),
		\label{eq:gt}
	\end{align}
\end{subequations}
where $\alpha = E, I$ represents the extracellular and intracellular ionic solution, $g^0_\alpha(t,t')$ and $\widetilde{g}^0_\alpha(t,t')$
are the usual quantum dissipative and fluctuational memory kernel in open quantum systems \cite{TuZhang2008,JTZY, LZ2012,YLZ,YZ2017}. They represent the original spectral structure of the ionic solution as well as ion back reactions between the ion channel and the ionic solution $\alpha$ without the external time-varying voltage (corresponds to $V_\alpha(t)=0$). The factor $M_\alpha(t,t')$ is defined as
\begin{align}
	M_\alpha(t,t') & \equiv  e^{-i[\phi_\alpha(t)-\phi_\alpha(t')]}, \notag \\
	& = \exp \Big\{\! - \!  \frac{i}{\hbar} \!  \int_{t'}^t \!  qV_\alpha(\tau)d\tau \Big\}  .
	\label{TTM}
\end{align}
It is determined by the voltage-associated flux phase difference in the ion tunnelings and characterizes the memory effect 
induced by the time-varying external voltage $V_\alpha(t)$.
	
This memory factor, $M_\alpha(t,t')$, modifies the ion channel spectral structure and gives rise to quantum coherence effect during the ion tunneling. It encodes the historical back reactions induced by the time-dependent external membrane potential. We therefore define these memory dynamics as active quantum memory. One can see that both Green functions $\bm{u}(t,t_0)$ and $\bm{v}(t,t')$ contain a time-convolution integral with the memory kernel matrix $\bm{g}(t,t')$ and $\widetilde{\bm{g}}(t,t')$, respectively, through Eq.~(\ref{eq:u}) and Eq.~(\ref{eq:FDT}). Consequently, via Eqs.~\eqref{eq:g} and \eqref{eq:gt}, the active quantum memory arising through the time-dependent voltage also affects the dynamics of these two Green functions. Thus, the above formulation provides an explicit description of active quantum memory in a single ion channel within a quantum memory circuit framework.

\section{Quantitative description of the active quantum memory}
	
\subsection{Transient ion transport dynamics in simple ion channel}
	
For a significant description of the active quantum memory, we consider the spectral density of the ionic environment with a wide band spectrum, namely the spectral density in the form of $J_\alpha(\epsilon) =\Gamma_\alpha$, where $\Gamma_\alpha$ is simply an energy-independent damping (dissipating) constant. A constant spectral density represents the ion tunneling from the ionic solution to the ion channel having equal probability for all different energy levels in the ionic solution. In other words, the background energy spectra of the ionic solution have a very flat distribution. Under such a simplification, the usual dissipative memory kernel in Eq.~(\ref{eq:gt}) becomes $g^0_\alpha(t,t') = \frac{\Gamma_\alpha}{2\hbar}\delta(t-t')$, so that the ionic environment itself does not generate non-Markovian memory dynamics.  The dissipation dynamics induced by the ionic solutions becomes a pure damping process. Meantime, the active quantum memory factor $M_\alpha(t,t')$ also makes no effect to the dissipation dynamics through $g_\alpha(t,t')$. This can be seen from the memory kernel in Eq.~(\ref{eq:gt}), $g_\alpha(t,t') = M_\alpha(t,t')  \frac{\Gamma_\alpha}{2\hbar} \delta(t-t') = \frac{\Gamma_\alpha}{2\hbar} \delta(t-t')$. Thus, the active quantum memory is fully manifested in the ion noise correlation through the fluctuational memory kernel $\widetilde{g}_\alpha(t,t')$ in the ionic transport current flowing the ion channel. 

To solve the ionic transport current, we will solve the retarded Green function $\bm{u}(t,t_0)$ first, i.e. the convolution differential equation of Eq.~(\ref{eq:u}), using the modified Laplace transform \cite{ZLXTN}. The modified Laplace transformation of Green function is reduced to
\begin{align}
	\bold{U}(z) = \Big[-iz\mathbb{I}_4 + \frac{i}{\hbar}\bm{\epsilon_s} + \bold{G}\Big]^{-1} = i\bm{A}^{-1}.
	\label{eq:Uz}
\end{align}
In the wideband spectrum $J_\alpha(\epsilon) =\Gamma_\alpha$, the dissipative memory kernel becomes memoryless and is given by a damping constant matrix after the transformation
\begin{align}
	\bm{G} =
	\begin{bmatrix}
		\frac{\Gamma_E}{2\hbar} & 0 & 0 & 0\\
		0 & 0 & 0 & 0\\
		0 & 0 & 0 & 0\\
		0 & 0 & 0 & \frac{\Gamma_I}{2\hbar}
	\end{bmatrix}.
\end{align} 
The matrix $\bm{A}(z)$ in Eq.~(\ref{eq:Uz}) is defined here for conveniently calculating the pole of $\bm{U}(z)$ by solving the following equation,
\begin{align}
	\det\{\bm{A}\} = 
	\begin{vmatrix}
		z-\frac{\epsilon_1}{\hbar}+i\frac{\Gamma_E}{2\hbar} & -\frac{\epsilon_{12}}{\hbar} & 0 & 0\\
		-\frac{\epsilon_{12}^*}{\hbar} & z-\frac{\epsilon_2}{\hbar} & -\frac{\epsilon_{23}}{\hbar} & 0\\
		0 & -\frac{\epsilon_{23}^*}{\hbar} & z-\frac{\epsilon_3}{\hbar} & -\frac{\epsilon_{34}}{\hbar}\\
		0 & 0 & -\frac{\epsilon_{34}^*}{\hbar} & z-\frac{\epsilon_4}{\hbar}+i\frac{\Gamma_I}{2\hbar}		
	\end{vmatrix}
	=0.
	\label{eq:det}
\end{align}
Explicitly, the solution of dissipation Green function $\bm{u}(t,t')$ is simply given by the inverse Laplace transformation method  
\begin{align}
	\bm{u}(t, t') = \frac{1}{2\pi}\sum_{z_p} i \bold{Z}(z_p) e^{-iz_p(t-t')}, 
	\label{eq:uwbl}
\end{align}
where for the clockwise contour integral,  
the matrix $\bm{Z}(z_p)$ is the residue of the pore $z_p$ in the the inverse Laplace transformation. 
It can be expressed as
\begin{align}
	\bm{Z}_{ij}(z_p) = \frac{-2\pi i[Adj(\bm{A})]_{ij}}{\frac{d}{dz}\{Det[\bm{A}(z)]\}}\biggr\rvert_{z=z_p} .
\end{align}

Thus, the ionic transport current determined by Eqs.~(\ref{eq:IE}) and (\ref{eq:II}) can be reduced to
\begin{align}
	I(t) =& I_E(t) - I_I(t) 
	\notag \\
	=& 2q \rm Re\Big\{\Big[\frac{\Gamma_E}{2\hbar}\bm{n}_{11}(t,t) - \frac{\Gamma_I}{2\hbar}\bm{n}_{44}(t,t)] \notag\\
	&- \!\! \int_{t_0}^t \!\! dt'  [\widetilde{g}_E(t,t')\bm{u}_{11}^\dag(t,t') - \widetilde{g}_I(t,t')\bm{u}_{44}^\dag(t,t')] \Big\} . \label{eq:current_wb}  
\end{align}
Obviously, the active quantum memory factors $M_\alpha(t,t')$ directly change the ion dynamics in the ion channel and then dominate the memory dynamics in the ion transport through the fluctuational memory kernel $\widetilde{g}_\alpha(t,t')$, in response to the external time-varying voltages exerted on the ion channel. 
Furthermore, the Fermi-Dirac distribution in the fluctuational memory kernel 
$\widetilde{g}_\alpha(t,t')$ can be rewritten in this form, 
\begin{align}
	f_\alpha(\epsilon) = \frac{1}{2}-\frac{1}{2} \tanh\Big[\frac{1}{2}\beta(\epsilon-\mu_\alpha)\Big].
\end{align}
Also, because $t>t'$ in Eq.~(\ref{eq:current_wb}),  the transformation in the frequency domain can be derived from 
the contour integral in the complex plane,
\begin{align}
	\int_{-\infty}^\infty &\frac{d\epsilon}{2\pi}\tanh\Big[\frac{1}{2}\beta(\epsilon-\mu_\alpha)\Big] e^{-\frac{i}{\hbar}\epsilon(t-t')} \notag\\
	&~~~~~~~~= -\frac{2i}{\beta}\sum_n e^{-\frac{i}{\hbar}(\mu_\alpha - \frac{i\pi}{\beta}(2n-1))(t-t')}.
\end{align}
Then the analytic form of the fluctuational memory kernel is obtained,
\begin{align}
	\widetilde{g}_\alpha(t,t')\! = & \frac{\Gamma_\alpha}{2\hbar}\delta(t-t') \notag\\ 
	&+ \frac{i\Gamma_\alpha}{\beta\hbar^2}M_\alpha(t,t')\!\!\sum_n e^{-\frac{i}{\hbar}(\mu_\alpha - \frac{i\pi}{\beta}(2n-1))(t-t')}.
	\label{eq:gt_wb}
\end{align}

Without loss of generality, we let the initial ion charges inside the ion channel become zero, $\bm{n}(t_0,t_0)=0$. Then, the matrix element of the total ion particle correlation inside the ion channel is given by 
\begin{widetext}
\begin{align}	
	\bm{n}_{ij}(t,t) =& \int_{t_0}^t \!\!\! d\tau \int_{t_0}^t \!\!\! d\tau' \bm{u}_{i1}(t,\tau)\widetilde{g}_E(\tau,\tau')\bm{u}^\dag_{1j}(t,\tau') + \bm{u}_{i4}(t,\tau) \widetilde{g}_I(\tau,\tau')\bm{u}^\dag_{4j}(t,\tau') \notag\\
	=& \frac{1}{(2\pi\hbar)^2}\sum_{z_pz_{p'}}\int_{t_0}^t\!\!d\tau\int_{t_0}^t\!\!d\tau'\int_{-\infty}^\infty\frac{d\epsilon}{2\pi}[\bm{Z}_{i1}(z_p)\bm{Z}_{1j}^\dag(z_{p'})\Gamma_Ef_E(\epsilon)M_E(\tau,\tau') \notag\\ 
	&~~~\qquad \qquad \qquad \qquad \qquad \qquad + \bm{Z}_{i4}(z_p)\bm{Z}_{4j}^\dag(z_{p'})\Gamma_If_I(\epsilon)M_I(\tau,\tau')]e^{-\frac{i}{\hbar}\epsilon(\tau-\tau')}e^{-i[z_p(t-\tau) - {z^*_{p'}}(t-\tau')]}.
	\label{particle number}
\end{align}
\end{widetext}
The expression for the total particle number of the ion channel, Eq.~\eqref{particle number}, reveals the connection between all binding sites and the ionic solutions through the matrix $\bm{Z}$. Although only the boundary sites ($i=1$ and $i=4$) couple directly to the extracellular and intracellular solutions in the Hamiltonian, coherent coupling among the binding sites enables all sites to interact with the environmental reservoirs. This is encoded in the nonzero matrix elements $\bm{Z}_{i1}$, $\bm{Z}_{1j}$, $\bm{Z}_{i4}$, and $\bm{Z}_{4j}$ in Eq.~\eqref{particle number}, where the indices $i,j$ can be any binding site within the channel. These matrix elements describe the effective dissipation strength of each site to the reservoirs, incorporating both the direct tunneling at the boundaries and the indirect coupling mediated by intersite transition. As a result, the ionic transport current passing through the ion channel is given by,
\begin{align}
		I(t) = & I_E(t) - I_I(t)    \notag \\
		= & \frac{2q}{\hbar} {\rm{Re}}\bigg\{\Big[\frac{\Gamma_E}{2}\bm{n}_{11}(t,t) - \frac{\Gamma_I}{2}\bm{n}_{44}(t,t)\Big] \notag\\
		& - \! \frac{i}{2\pi\hbar} \!\! \int_{-\infty}^\infty \!\! \frac{d\epsilon}{2\pi} \!\! \int_{t_0}^t \!\!\! dt'\sum_{z_p}\big[\bm{Z}_{11}(z_p)\Gamma_Ef_E(\epsilon)M_E(t',t) \notag \\ 
		& \qquad \!- \! \bm{Z}_{44}(z_p)\Gamma_If_I(\epsilon)M_I(t',t)\big] e^{(-iz_p+\frac{i}{\hbar}\epsilon)(t-t')}\bigg\}.  
\label{eq:current_wb_exact}
\end{align}
The responses of the ion transport current to the external time-varying voltage describe the detailed active quantum memory dynamics in the ion channel.

\subsection{Active quantum memory induced by a periodic oscillating bias potential}

The active quantum memory factor $M_\alpha(t,t')$ defined in Eq.~(\ref{TTM}) is closely correlated with the quantum coherence in the ion channel dynamics, responding to the time-varying external voltage $V_\alpha(t)$. In order to find the general rules of the active quantum memory, we now apply a symmetrical sinusoidal varying voltage to the ion solutions as the simplest signal exerted on the membrane. These voltage functions are superimposed on the chemical potentials $\mu_I$ and $\mu_E$:
\begin{subequations}
	\begin{align}
		V_I(t) = \frac{1}{2}V(t),& ~V_E(t) = - \frac{1}{2}V(t) ,  \label{eq:eachbias}\\
		V(t) = V_I(t) - &V_E(t) = V_d\sin(\omega_d t).
		\label{eq:bias}
	\end{align}
\end{subequations}
The time-varying voltage in Eq.~\eqref{eq:bias} represents the time varying membrane potential difference between the intracellular and extracellular ion solutions surrounding the membrane. In the initial equilibrium state for the ion solutions, we assume $\mu_I =\mu_E$ for focusing on the time-dependent voltage. Then the active quantum memory factor of Eq.~(\ref{TTM}) becomes  
\begin{align}
	M_{I,E}(t,t') = \exp\Big\{\pm i \frac{2\pi\Phi_M}{\Phi_0} \big[\cos\omega_dt-\cos\omega_dt'\big] \Big\}, 
\end{align}
where $\Phi_M=V_d/2\omega_d $ is the amplitude of the oscillating magnetic flux 
and $\Phi_0=h/q$ is the magnetic flux quanta, as mentioned before. 
The plus and minus signs represent respectively the external voltage $V_{I,E}(t)$ exerting on the intracellular and extracellular ionic solution $\alpha=I,E$ on the membrane surface. 
Thus, the fluctuational memory kernel given by Eq.~\eqref{eq:gt_wb} under the wide-band limit spectral density can be expanded in terms of the Bessel function of the first kind, $J_m(x)$, as follows, 
\begin{align}
	\begin{split}
		\widetilde{g}_{I,E} &(t, t')  =  \frac{\Gamma_\alpha}{2\hbar}\delta(t-t')\\
		&+\frac{i\Gamma_\alpha}{\beta\hbar^2}\sum_{m=-\infty}^\infty\!\sum_{m'=-\infty}^\infty \!\! J_m(\phi_M) J_{m'}(\phi_M)\\
		& \times i^{\pm(m-m')}e^{\pm i\omega_c(mt-m't')} e^{-\frac{i}{\hbar}(\mu_\alpha-\frac{i\pi}{\beta}(2n-1))(t-t')},
		\label{eq:gt_sinV}
	\end{split}
\end{align}
where 
\begin{align}
	\phi_M = 2\pi \frac{\Phi_M}{\Phi_0} =\frac{V_d}{2\omega_d}\frac{q}{\hbar}    \label{amqp}
\end{align}
denotes the effective magnetic flux phase strength associated with the active driving voltage or equivalently the effective 
driving magnetic flux. The summations of $m$ and $m'$ modulate the oscillation frequency, generating the sideband current in the ion channel. Although the term "sideband current" is borrowed from signal processing, the physical mechanism differs fundamentally from classical amplitude modulation. In radio communications, sidebands arise from external modulation of carrier signals. While in the ion channel system, sidebands arise from quantum coherence in ion tunneling. It is the time-dependent sinusoidal voltage that induces an effective magnetic flux phase in the memory kernel Eq.~\eqref{eq:gt_sinV}, generating multiple components at frequencies different from the driving frequency. These additional frequency components are quantum mechanical in origin, reflecting coherent superposition in the tunneling process.

Now we can discuss the detailed ion transport dynamics through the quantum mechanical multi-level ion channel. For simplifying the ion channel structure without the loss of generality, we set $\epsilon_1 = \epsilon_4$ and $\epsilon_2 = \epsilon_3$. The intersite transitions are uniformly defined as $J$. And the coupling strengths to the intracellular and the extracellular ionic environments are set the same as $\Gamma_I = \Gamma_E = \frac{\Gamma}{2}$ for considering the selectivity filter symmetrically. Then the poles of $\bm{U}(z)$ from Eq.~(\ref{eq:det}) can be easily solved
\begin{align}
	z_{p1} = &\frac{1}{2\hbar}\Big[\epsilon_1\! +\! \epsilon_2\! -\! J \!-i\frac{\Gamma}{4} -\! \sqrt{(\epsilon_1\! - \epsilon_2\! +\! J\! - i\frac{\Gamma}{4})^2 + 4J^2 }\Big], \notag\\
	z_{p2} = &\frac{1}{2\hbar}\Big[\epsilon_1\! +\! \epsilon_2\! -\! J \!-i\frac{\Gamma}{4} +\! \sqrt{(\epsilon_1\! - \epsilon_2\! +\! J\! - i\frac{\Gamma}{4})^2 + 4J^2 }\Big], \notag\\ 
	z_{p3} = &\frac{1}{2\hbar}\Big[\epsilon_1\! + \!\epsilon_2\! +\! J \!-i\frac{\Gamma}{4} +\!\sqrt{(\epsilon_1\! - \epsilon_2\! -\! J\! - i\frac{\Gamma}{4})^2 + 4J^2}\Big], \notag\\
	z_{p4} = &\frac{1}{2\hbar}\Big[\epsilon_1\! + \!\epsilon_2\! +\! J \!-i\frac{\Gamma}{4} -\! \sqrt{(\epsilon_1\! - \epsilon_2\! -\! J\! - i\frac{\Gamma}{4})^2 + 4J^2}\Big].
	\label{eq:pole}
\end{align}
These poles solved from the retarded Green function $\bm{u}(t,t_0)$ indicate that the intersite couplings between the original four sites in the ion Kv1.2 channel and the coupling of the first and the last sites to the extracellular and intracellular ion solutions, respectively, make the original four on-site energy states become four renormalized states. Usually, the quantum transport is driven by the hopping processes. In the multi-level system, such as the Hamiltonian in Eq.~\eqref{eq:Kv1.2 Hamiltonian}, the hopping processes between sites generate four eigen modes. Each eigen modes is a superposition of the original binding sites (four on-site energy states). Specifically, the eigen modes arise from the diagonalization of the system Hamiltonian; they can be obtained by diagonalizing Eq.~\eqref{bs}. Furthermore, due to the first and the last binding sites coupled with the ionic solutional environment, the coupling strength affects the ion transport processes, generating the original eigen mode to turn into the non-zero decay resonant states. Consequently, the four renormalized states are not the bound states; they are usually called resonant states with non-zero decay width.  The real parts of these poles correspond to the renormalized energies of these resonant states, $\epsilon'_i = \hbar {\rm Re}\{z_{p_i}\}, ~i=1,2,3,4$, and the imaginary parts represent the damping (decay) constants of these resonant states. These basic quantum mechanical solutions cannot be obtained and described in the classical picture.

\begin{figure*}
	\centering
	\includegraphics[width=1\linewidth]{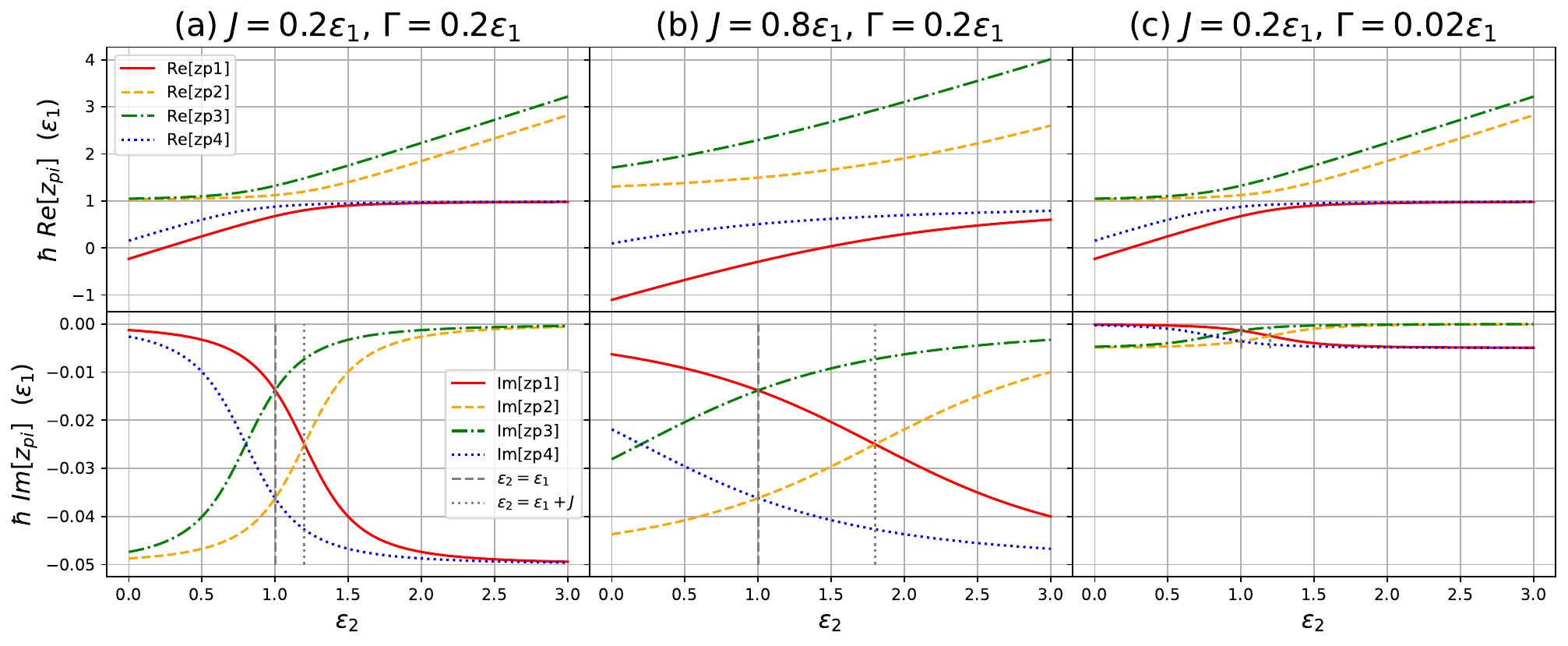}
	\caption{(color online). Changes of four poles in the retarded Green's function as a function of on-site energy level $\epsilon_2$ varied from 0 to $3\epsilon_1$. The upper and lower panels show the real and imaginary parts of the poles, respectively. The parameters are taken as (a) $J=0.2\epsilon_1$, $\Gamma = 0.2\epsilon_1$; (b) $J=0.8\epsilon_1$, $\Gamma = 0.2\epsilon_1$; and (c) $J=0.2\epsilon_1$, $\Gamma = 0.02\epsilon_1$. The first on-site energy level $\epsilon_1$ is taken as unit.}
	\label{fig3:pole}
\end{figure*}

Figure~\ref{fig3:pole} shows the behaviour of the four poles in Eq.~\eqref{eq:pole} as a function of energy level $\epsilon_2$ 
changes in units of $\epsilon_1$. The upper and lower panels show the real and imaginary parts 
of the four poles, respectively. We present three different parameter sets: (a) $J=0.2\epsilon_1$, 
$\Gamma=0.2\epsilon_1$ (moderate interdot tunneling), (b) $J=0.8\epsilon_1$, 
$\Gamma=0.2\epsilon_1$ (strong tunneling), and (c) $J=0.2\epsilon_1$, $\Gamma=0.02\epsilon_1$ 
(weak coupling). In all cases, we vary $\epsilon_2$ from 0 to $3 \epsilon_1$
The real parts exhibit similar patterns in all three parameter sets. Taking Fig.~\ref{fig3:pole}(a) for an example, 
as $\epsilon_2$ becomes larger, two of the real parts approach $\epsilon_2\pm J$, while the other two real parts ($z_{p1}$ and 
$z_{p4}$) approach $\epsilon_1$. When the transition energy is larger [see Fig.~\ref{fig3:pole}(b) with $J=0.8\epsilon_1$], 
the separation between the real parts widens, indicating enhanced level splitting. Weakening the reservoir coupling $\Gamma$ primarily reduces the imaginary parts (the damping or decay rate) while leaving the real parts of poles (the renormalized 
level) nearly unchanged. The imaginary parts are sensitively determined by the reservoir coupling strength $\Gamma$
[see Fig.~\ref{fig3:pole}(c) with $\Gamma=0.02\epsilon_1$], as we expected. 
Notably, when the system has the fully degeneracies ($\epsilon_2=\epsilon_1$),​ it leads to pairwise equal decay rate. Similarly, 
at $\epsilon_2=\epsilon_1+J$, two of the decay rates become identical. These crossings reflect the underlying symmetries 
of this four-level model.

After analyzing the properties of the four resonant energy states through the pole structure of the retarded Green’s function, the physical quantities relevant to ion transport can be derived more explicitly. Given the explicit form of the quantum memory kernel $\widetilde{g}_\alpha(t,t')$ in Eq.~\eqref{eq:gt_sinV}, 
the ion occupation number in the channel at an arbitrary time $t$ can be obtained,
\begin{widetext}
	\begin{align}
		\begin{split}
			\bold{n}_{ij}(t,t) = &\frac{1}{(2\pi\hbar)^2} \!\! \sum_{z_pz_{p'}} \!\! \Bigg\{ \! \frac{1-e^{-i(z_p-z_{p'}^*)t}}{i(z_p - z_{p'}^*)}\Big[\frac{\hbar\Gamma_E}{2}\bm{Z}_{i1}(z_p)\bm{Z}_{1j}^*(z_{p'}) + \frac{\hbar\Gamma_I}{2}\bm{Z}_{i4}(z_p)\bm{Z}_{4j}^*(z_{p'}) \Big] \!+\!\!\!\!\! \sum_{m, m'=-\infty}^\infty 
			\! \sum_{n=1}^\infty J_m(\phi_M)J_{m'}(\phi_M)\\ &\times\bigg[ \bm{Z}_{i1}(z_p)\bm{Z}_{1j}^\dag(z_{p'})\frac{i\Gamma_E}{\beta}i^{-(m-m')}\bigg(\frac{\frac{e^{-i\omega_d(m-m')t} - e^{-i(z_p - z_{p'}^*)t}}{-i(\omega_dm - \omega_dm' - z_p + z_{p'}^*)} - \frac{e^{i(z_{p'}^* - \frac{\mu_E}{\hbar} - \omega_dm + \frac{i\pi}{\hbar\beta}(2n-1))t} - e^{-i(z_p - z_{p'}^*)t}}{i(z_p - \frac{\mu_E}{\hbar} - \omega_dm + \frac{i\pi}{\hbar\beta}(2n-1))}}{i(\omega_dm' - z_{p'}^* + \frac{\mu_E}{\hbar} - \frac{i\pi}{\hbar\beta}(2n-1))}\\
			& \qquad \qquad \qquad \qquad \qquad \qquad ~~~~ -\frac{(\frac{e^{-i\omega_d(m-m')t} - e^{-i(z_p - z_{p'}^*)t}}{-i(\omega_dm - \omega_dm' - z_p + z_{p'}^*)} - \frac{e^{-i(z_{p}-\frac{\mu_E}{\hbar}-\omega_dm'-\frac{i\pi}{\hbar\beta}(2n-1))t} - e^{-i(z_p-z_{p'}^*)t}}{-i(z_{p'}^* - \frac{\mu_E}{\hbar} - \omega_d{m'} - \frac{i\pi}{\hbar\beta}(2n-1))})}{-i(\omega_dm - z_p + \frac{\mu_E}{\hbar} + \frac{i\pi}{\hbar\beta}(2n-1))}\bigg)\\
			&+ \bm{Z}_{i4}(z_p)\bm{Z}_{4j}^\dag(z_{p'})\frac{i\Gamma_I}{\beta}i^{(m-m')}\bigg(\frac{\frac{e^{i\omega_d(m-m')t} - e^{-i(z_p - z_{p'}^*)t}}{i(\omega_dm - \omega_dm' + z_p - z_{p'}^*)} - \frac{e^{i(z_{p'}^* - \frac{\mu_I}{\hbar} + \omega_dm + \frac{i\pi}{\hbar\beta}(2n-1))t} - e^{-i(z_p-z_{p'}^*)t}}{i(z_p - \frac{\mu_I}{\hbar} + \omega_dm + \frac{i\pi}{\hbar\beta}(2n-1))}}{-i(\omega_dm'+z_{p'}^* - \frac{\mu_I}{\hbar} + \frac{i\pi}{\hbar\beta}(2n-1))}\\
			&\qquad \qquad \qquad \qquad \qquad \qquad ~~~~-\frac{\frac{e^{i\omega_d(m-m')t} - e^{-i(z_p-z_{p'}^*)t}}{i(\omega_dm - \omega_dm' + z_p - z_{p'}^*)} - \frac{e^{-i(z_p - \frac{\mu_I}{\hbar} + \omega_dm' - \frac{i\pi}{\hbar\beta}(2n-1))t} - e^{-i(z_p-z_{p'}^*)t}}{-i(z_{p'}^* - \frac{\mu_I}{\hbar} + \omega_dm' - \frac{i\pi}{\hbar\beta}(2n-1))}}{i(\omega_dm + z_p - \frac{\mu_I}{\hbar} - \frac{i\pi}{\hbar\beta}(2n-1))}\bigg)\bigg]\Bigg\}.
		\end{split}
	\end{align}
\end{widetext}
Then the ion transient transport current can be calculated directly through Eq.~(\ref{eq:current_wb_exact}),
\begin{widetext}
	\begin{align}
			I(t) = &\frac{2q}{\hbar} {\rm{Re}}\Bigg\{\Big[\frac{\Gamma_E}{2}\bm{n}_{11}(t,t) - \frac{\Gamma_I}{2}\bm{n}_{44}(t,t)\Big] - \frac{i}{8\pi}\sum_{z_p}(\Gamma_E\bm{Z}_{11}(z_p) - \Gamma_I\bm{Z}_{44}(z_p)) \notag\\
			&-\frac{1}{2\pi\beta\hbar} \!\!\! \sum_{m,m'=-\infty}^\infty 
			\!\sum_{n=1}^\infty\!\sum_{{z_p}}\!J_m(\phi_M)J_{m'}(\phi_M) \! \bigg[\Gamma_E\bm{Z}_{11}(z_p)i^{-(m-m')}\frac{e^{-i(m-m')\omega_dt} - e^{-i(z_p - \frac{\mu_E}{\hbar} - \frac{i\pi}{\beta\hbar}(2n-1)-m'\omega_d)t}}{i(z_p - \frac{\mu_E}{\hbar} - \frac{i\pi}{\hbar\beta}(2n-1) - m\omega_d)} \notag\\
			& \qquad \qquad \qquad \qquad \qquad \qquad \qquad \qquad ~~ - \! \Gamma_I\bm{Z}_{44}(z_p)i^{(m-m')}
			\frac{e^{i(m-m')\omega_dt} - e^{-i(z_p-\frac{\mu_I}{\hbar}-\frac{i\pi}{\beta\hbar}(2n-1) + m'\omega_d)t}}{i(z_p - \frac{\mu_I}{\hbar} - \frac{i\pi}{\hbar\beta}(2n-1) + m\omega_d)}\bigg]\Bigg\}.
			\label{eq:WBL_current}
	\end{align}
\end{widetext}
The quantum coherence effect induced by the time-varying voltage is demonstrated with an oscillatory transport current through the ionic tunneling. The transport current depends on three key parameters: the renormalized resonance energy levels $\epsilon_i'$, the chemical potentials $\mu_\alpha$, and the driving voltage amplitude $V_d$. The chemical potentials specify the energy distribution of occupied states in two ionic leads. In particular, when both chemical potentials align with a specific renormalized energy level, the resonant tunneling is maximized. The strongest resonance enhances the current amplitude and produces a pronounced sideband structure, with ions mainly tunnel through the resonant level.

\begin{figure}
	\centering
	\includegraphics[width=1\linewidth]{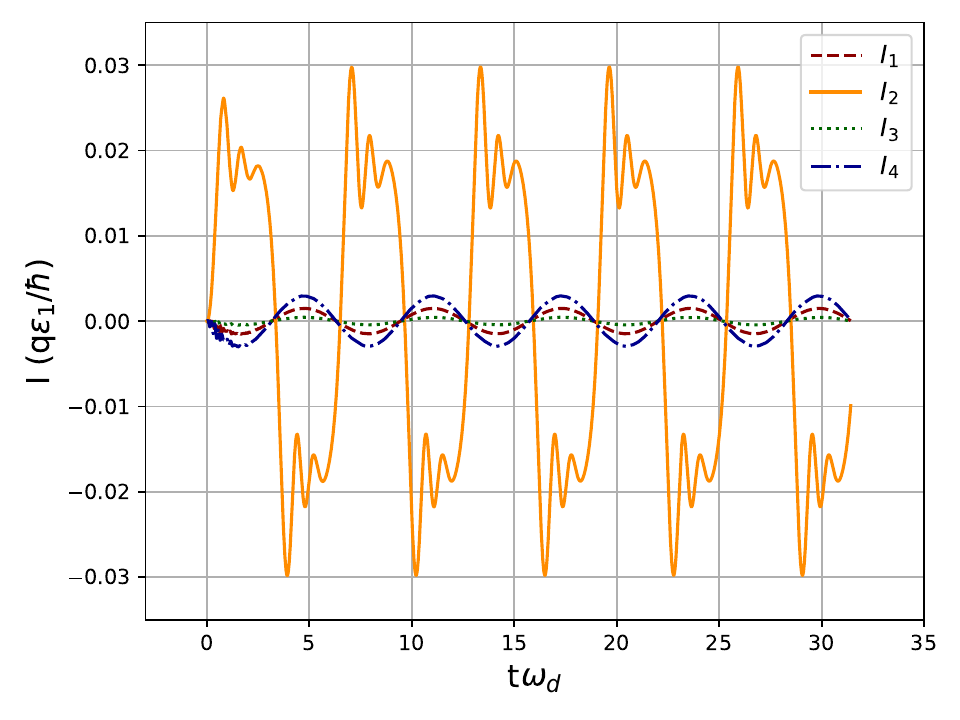}
	\caption{(color online). Ion transport currents contributed from each renormalized energy level. Current $I_i$ represents the current from the $i_{th}$ renormalized level with energy $\epsilon_i'$. In the selected parameter setup, the current from the second renormalized level exhibits strong sideband structure, while contributions from the other three renormalized levels are negligible. The parameters are chosen as $\epsilon_1=\epsilon_4$, $\epsilon_2=\epsilon_3 = 2.5\epsilon_1$ and the transition energy $J=1.2\epsilon_1$. The chemical potential is set as $\mu=2.36\epsilon_1$. The driving voltage is given by Eq.~\eqref{eq:bias} with amplitude $qV_d = 0.8\epsilon_1$ and frequency $\hbar\omega_d = 0.05\epsilon_1$. Other parameters are taken as: the temperature $kT = 0.001\epsilon_1 $, the couplings of both intracellular and extracellular ionic leads $\Gamma_I = \Gamma_E = 0.1\epsilon_1$. All parameters are set in terms of the energy unit $\epsilon_1$.}
	\label{fig4:current_from_level}
\end{figure}

Here, we first give an example of the ion transport current in Fig.~\ref{fig4:current_from_level}, with $\epsilon_1=\epsilon_4$, $\epsilon_2=\epsilon_3 = 2.5\epsilon_1$ and the transition energy $J=1.2\epsilon_1$. For this parameter setup, the poles of the retarded Green's function 
are given by
\begin{align}
	\hbar z_{p_1} &= (-0.059 - 0.028i) \epsilon_1, \notag\\
	\hbar z_{p_2} &= (2.359 - 0.022i) \epsilon_1, \notag\\
	\hbar z_{p_3} &= (4.156 - 0.006i) \epsilon_1, \notag\\
	\hbar z_{p_4} &= (0.544 - 0.043i) \epsilon_1, 
\end{align}
where the real parts represent the renormalized energy levels $\epsilon_i'$, each of them is a superposition of the original four site states in the ion channel. To focus on the time-varying driving voltage effects, we have taken the chemical potentials of the two ion solutions equal: $\mu_I = \mu_E = \mu$. The chemical potential is set as $\mu = 2.36 \epsilon_1$, closed to the second energy state $\epsilon_2' = 2.359 \epsilon_1$. The result shows that the tunneling through the second renormalized energy state has the largest transport current and demonstrates the significant sideband structure, while the remaining currents through the other three renormalized energy levels are negligible. This shows that the ion transport is dominated by the resonant tunneling level.

\begin{figure*}
	\centering
	\includegraphics[width=0.9\linewidth]{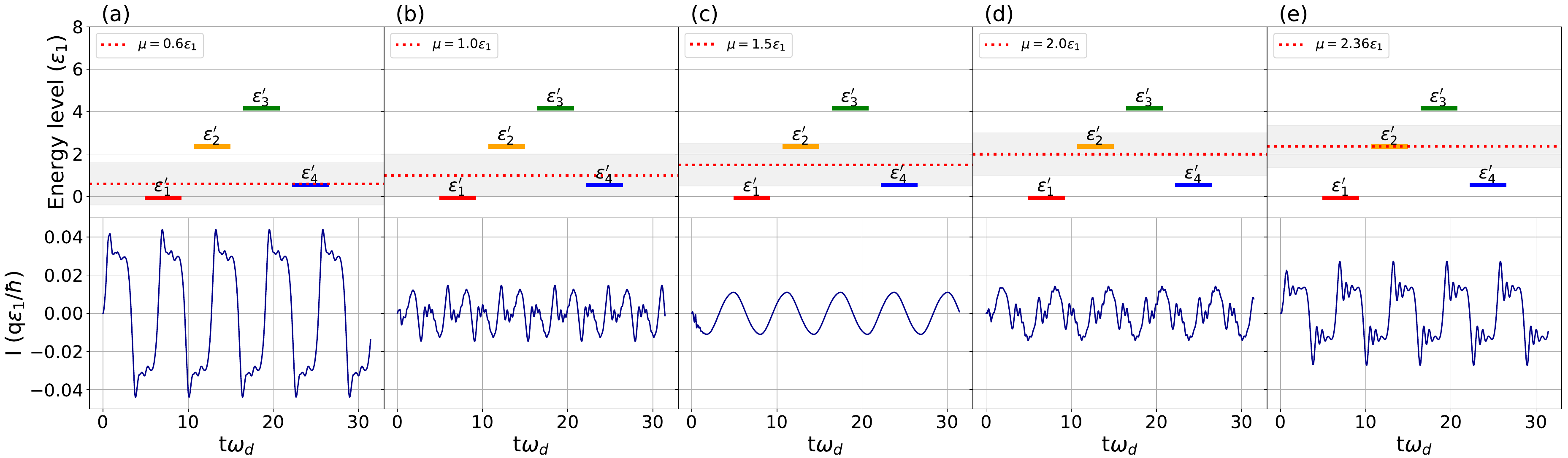}
	\caption{(color online). The ion transport with different chemical potentials: (a) $\mu=0.6\epsilon_1$, (b) $\mu=1\epsilon_1$, (c) $\mu=1.5\epsilon_1$, (d) $\mu=2\epsilon_1$, (e) $\mu=2.36\epsilon_1$. The time-varying voltages are given in Eq.~\eqref{eq:bias} with parameters $qV_d = 1\epsilon_1$ and $\hbar\omega_d = 0.05\epsilon_1$. The upper panel shows the renormalized energy levels (with  the colored bars represent the renormalized energy levels $\epsilon_1',...\epsilon_4'$). The horizontal dashed line marks the chemical potential $\mu$, and the shaded region [$\mu-qV_d, \mu+qV_d$] indicates the energy window of for the resonant tunneling. The lower panel  is the calculated ion current. The on-site binding site energies are taken as $\epsilon_1 = \epsilon_4$ and $\epsilon_2 = \epsilon_3 = 2.5\epsilon_1$. The transition energy is set as $J = 1.2\epsilon_1$. Other parameters are taken as: the temperature $kT = 0.001\epsilon_1 $, the couplings of both intracellular and extracellular ionic leads $\Gamma_I = \Gamma_E = 0.1\epsilon_1$. All parameters are gien in terms of the energy unit $\epsilon_1$.}
	\label{fig5:adj_chemical_potential}
\end{figure*}
\begin{figure*}
	\centering
	\includegraphics[width=0.6\linewidth]{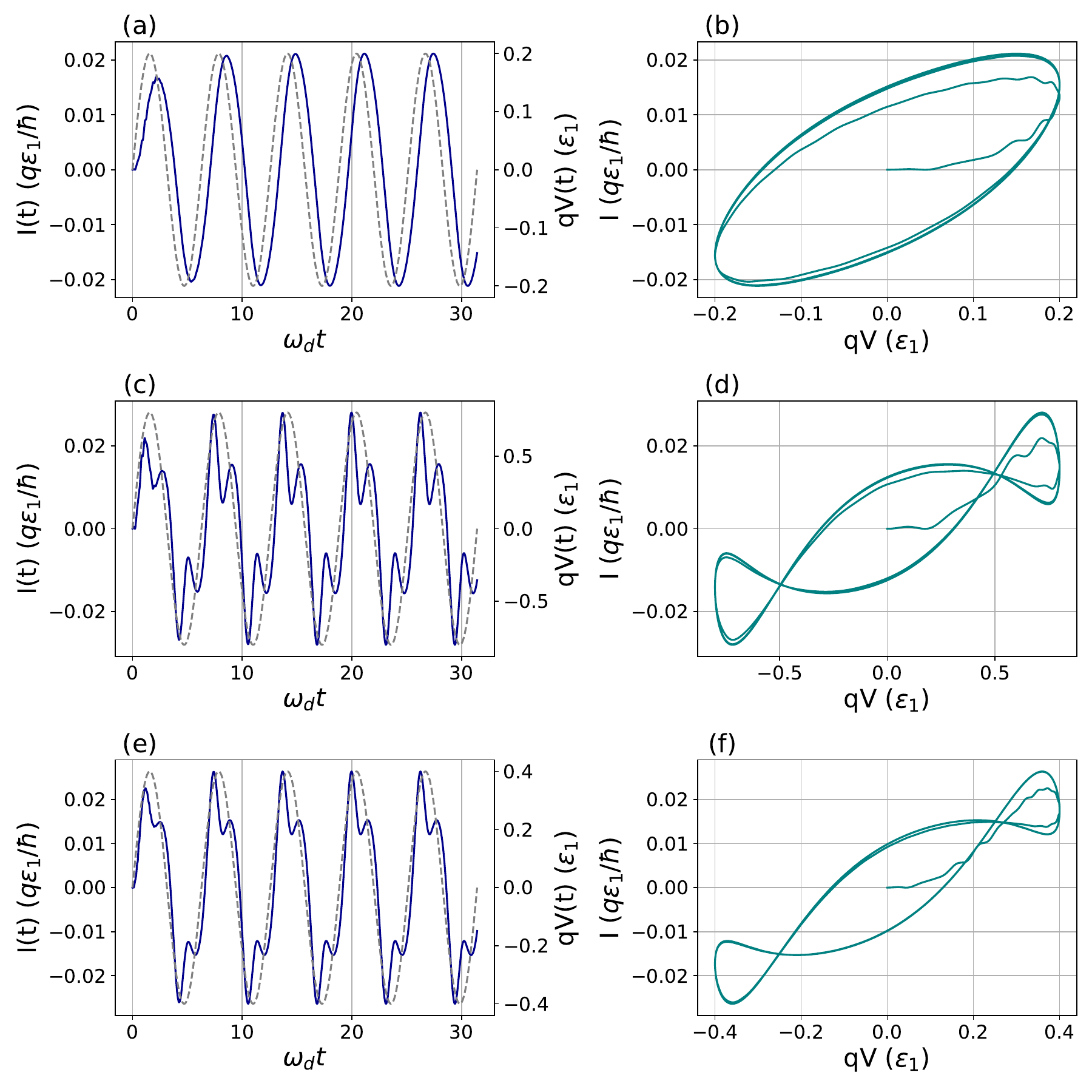}
	\caption{(color online). The currents and the current-voltage characteristics (I-V curve) derived from Eq.~\eqref{eq:WBL_current}. The left panel shows the transient ion channel currents in response to the time-varying bias voltage (dashed line). The right panel is the corresponding I-V curve. The parameters are set for (a)-(b) $qV_d = 0.2\epsilon_1$, $\hbar \omega_d = 0.1\epsilon_1$ ($\phi_M = 1$); (c)-(d) $qV_d = 0.8\epsilon_1$, $\hbar \omega_d = 0.1\epsilon_1$ ($\phi_M = 4$); and (e)-(f) $qV_d = 0.4\epsilon_1$, $\hbar \omega_d = 0.05\epsilon_1$ ($\phi_M = 4$). The temperature is $kT = 0.001\epsilon_1 $, and the couplings of both intracellular and extracellular ionic leads are $\Gamma_I = \Gamma_E = 0.1\epsilon_1$. The parameters of the binding site energy are $\epsilon_1 = \epsilon_4$ and $\epsilon_2 = \epsilon_3 = 2.5\epsilon_1$. The transition energy is set as $J = 1.2\epsilon_1$. The chemical potentials are equal $\mu_E = \mu_I = 2.36\epsilon_1$. All parameters are set in terms of the energy unit $\epsilon_1$.}
	\label{fig6:transient current}
\end{figure*}

The above example of Fig.~\ref{fig4:current_from_level} shows that, other than the resonant state aligning with the chemical potential, only a few ions tunnel through the other three resonant states. More specifically, the energy distance between the chemical potentials and the renormalized levels influences the tunneling strength. To illustrate this tunneling behaviour, we vary $\mu$ across different renormalized energy levels; see Fig.~\ref{fig5:adj_chemical_potential}, keeping all other parameters fixed. The resonance occurs when the renormalized levels lie within the time-varying voltage window range within [$\mu-qV_d$, $\mu+qV_d$]. The strongest resonant current occurs when $\mu$ aligns with a renormalized energy level [see Fig.~\ref{fig5:adj_chemical_potential}(a), (e)], yielding pronounced and significant sidebands with larger current amplitude. When $\mu$ deviates from the renormalized energy level [see Fig.~\ref{fig5:adj_chemical_potential}(b), (d)], the resonant tunneling becomes weaker, and the current amplitudes are significantly decreased. As the energy window lies between two renormalized energy levels [see Fig.~\ref{fig5:adj_chemical_potential}(c)], the resonant tunneling is too weak to show the multiple oscillatory structure, so the current follows in a simple sinusoidal form.
We find that when the chemical potential $\mu$ reaches to one renormalized energy level $\epsilon_i'$, and $qV_d$ is smaller than the nearest-neighbor level spacing of the matched level ($qV_d \ll |\epsilon_{i}'-\epsilon_{j}'|$), the ion current displays a strong sideband structure characterizing the strongest resonance [see Fig.~\ref{fig5:adj_chemical_potential}(e)]. The corresponding sideband structures are also characteristic signatures of photon-assisted resonant tunneling, as observed in single-level resonant tunneling systems \cite{WJM,JWM,JTZY}. Importantly, each renormalized level $\epsilon_i'$ involves a quantum superposition of original four on-site binding sites $\epsilon_1$, $\epsilon_2$, $\epsilon_3$, $\epsilon_4$. Therefore, even when the ion transport appears to proceed through a single renormalized energy level, it intrinsically involves the contributions from all four on-site energy states in Fig.~\ref{fig2:Kv1.2 ion channel}. This finding cannot be seen in any classical picture.

\begin{figure*}
	\centering
	\includegraphics[width=0.98\textwidth]{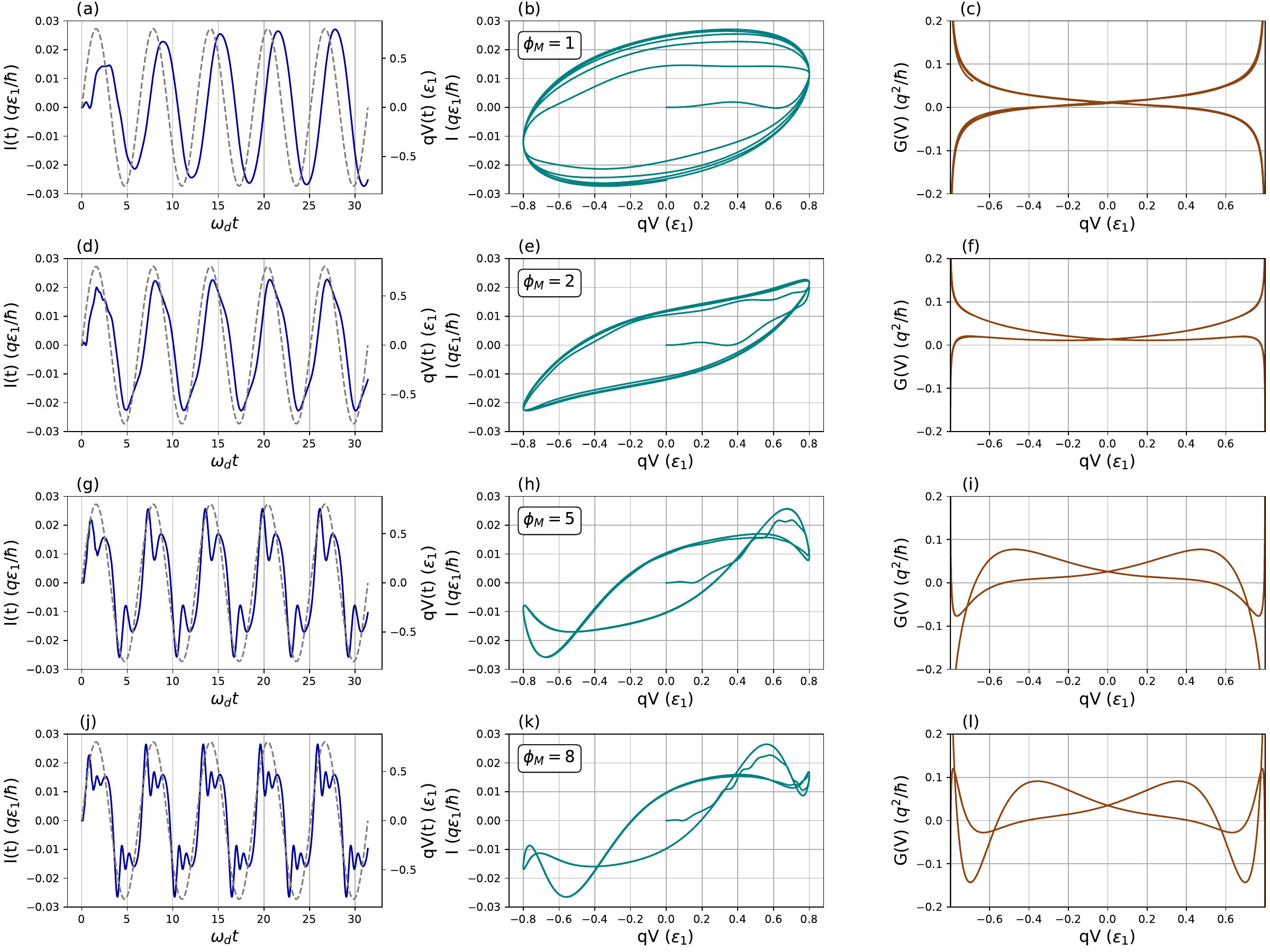}
	\caption{(color online) Active quantum memory in the ion transport in ion channel. The ion current (dark-blue line) changed along with the sinusoidal bias voltage(dashed line) is presented in the left panel, the corresponding I-V characteristics (green line) in the middle panel, and the steady-state differential conductance (brown line) in the right panel. Four flux phase strengths are set with (a)-(c) $\phi_M = 1$ ($\hbar\omega_d = 0.4\epsilon_1$); (d)-(f) $\phi_M = 2$ ($\hbar\omega_d = 0.2\epsilon_1$); (g)-(i) $\phi_M = 5$ ($\hbar\omega_d = 0.08\epsilon_1$); (j)-(l) $\phi_M = 8$ ($\hbar\omega_d = 0.05\epsilon_1$). The driving voltage amplitude is fixed to $qV_d=0.8\epsilon_1$. Other parameters are set as: the temperature $kT = 0.001\epsilon_1$, the couplings $\Gamma_L = \Gamma_R = 0.1\epsilon_1$, the binding-site energies $\epsilon_1 = \epsilon_4$, $\epsilon_2 = \epsilon_3 = 2.5\epsilon_1$ and transition energy $J=1.2\epsilon_1$. The initial ion occupation number in the ion channel $n(t_0)=0$. The chemical potentials are $\mu_I = \mu_E = 2.36\epsilon_1$.}
	\label{fig7:std_current_r}
\end{figure*}

Now, we can examine how the sideband structure depends on the time-varying driving voltage amplitude $qV_d$ and the frequency $\hbar\omega_d$. The sideband oscillations arise from the dynamical memory effects encoded in the magnetic flux phase $\phi_M$ in Eq.~\eqref{amqp}, where the parameter characterizes the active quantum memory. For sinusoidal driving, $\phi_M$ depends on the ratio $qV_d / 2\hbar\omega_d$, which combines the amplitude and frequency into a single control parameter to the active quantum memory. Consequently, increasing $qV_d$ or decreasing $\hbar\omega_d$ produces equivalent effects on the sideband dynamics. These time-varying voltage-induced memory features are most clearly revealed in the current-voltage (I-V) characteristics, which display hysteresis loops whose complexity increases with varying $\phi_M$. In the following example, we fix the chemical potential at $\mu=2.36\epsilon_1$. Figure~\ref{fig6:transient current} shows three representative cases with different driving parameters. For all cases, we initialize the system with zero ion occupation, $n(t_0) = 0$. Figure~\ref{fig6:transient current}(a) is the ion current for $\hbar\omega_d = 0.1\epsilon_1$ and $qV_d = 0.2\epsilon_1$ ($\phi_M=1$). After two driving cycles, the current reaches steady state with simple sinusoidal response with a phase shift relative to the driving voltage. The corresponding I-V curve [see Fig.~\ref{fig6:transient current}(b)] forms an ellipse without sidebands. This result is the same as the I-V curve of the usual classical RLC circuit formed by the passive electronic components.

However, when the amplitude increases to $qV_d=0.8\epsilon_1$ and keeps the frequency $\hbar\omega_d=0.1\epsilon_1$ ($\phi_M=4$) [see Fig.~\ref{fig6:transient current}(c)], the memory effect associated with the magnetic flux phase strengthens. As a result, the current oscillates beyond a simple sinusoidal form, showing the sideband structure, and the I-V characteristic forms non-zero crossing points, as shown in Fig.~\ref{fig6:transient current}(d). Alternatively, reducing the frequency to $\hbar\omega_d=0.05\epsilon_1$, while keeping $qV_d = 0.2\epsilon_1$ (yielding $\phi_M=4$), produces qualitatively similar features, as shown in Fig.~\ref{fig6:transient current}(e), (f). Overall, increasing the driving amplitude or decreasing the frequency enhances the active memory effect, as evidenced by stronger sideband oscillations and the increasingly complex I-V hysteresis. All these features scale with $\phi_M$, confirming its role as the key control parameter. This phenomenon results from quantum coherence governed by the active quantum memory factor $M(t,t')$ in Eq.~(\ref{TTM}). The active quantum memory governs the emergence of sideband structure and hysteresis in ion channel transport. In the next, we will quantify these memory effects through a systematic analysis of the hysteresis crossing points.

As shown in the previous examples [see Fig.~\ref{fig6:transient current}],  the ion transport current reaches the steady state within approximately two to three driving periods. In Eq.~\eqref{eq:WBL_current}, the magnetic flux phase strength $\phi_M$ in the Bessel function characterizes the quantum coherence induced by the external time-varying voltage through the effective magnetic flux, as seen in the definition given by Eq.~(\ref{amqp}). Figure~\ref{fig6:transient current} shows that the complexity of sideband dynamics increases with relatively larger $\phi_M$. Consequently, $\phi_M$ quantifies the active quantum memory, characterizing the influence of the driving bias voltage in terms of the effective magnetic flux.
We now systematically vary $\phi_M$ to show its effect on the ion current and the I-V curve. For small $\phi_M$, the memory effects are weak. For example, at $\phi_M=1$ [see Fig.~\ref{fig7:std_current_r}(a)], the ion current shows a simple sinusoidal form with a phase shift relative to the driving voltage, analogous to the previous example as shown in Fig.~\ref{fig6:transient current}(a). The I-V curve forms a simple ellipse without crossings [see Fig.~\ref{fig7:std_current_r}(b)]. When the flux phase strength increases to $\phi_M=2$, the memory effect strengthens, distorting the I-V curve from the elliptical shape [see Fig.~\ref{fig7:std_current_r}(d), (e)]. For larger $\phi_M$, pronounced sidebands appear, and the non-zero crossing points develop in the I-V hysteresis loop. At $\phi_M = 5$ [see Fig.~\ref{fig7:std_current_r}(h)], the I-V curve exhibits two crossing points, and at $\phi_M = 8$ [see Fig.~\ref{fig7:std_current_r}(k)], the number of crossing points increases to four. We therefore define the active quantum memory strength as the number of hysteresis crossing points.

To further characterize the sideband dynamics, we examine the differential conductance 
\begin{align}
	G(t) = {dI(t)}/{dV_b(t)}.  \label{dcond}
\end{align} 
The right panel of Fig.~\ref{fig7:std_current_r} shows the $G-V$ characteristics, where the differential conductance oscillates between plus and minus infinity. These divergences are a sort of Fano resonance. The divergent conductances arise at the tuning points of the I-V trajectory.  For the weak active quantum memory ($\phi_M = 1$), $G-V$ curve only crosses itself once at $qV_b=0$, as shown by Fig.~\ref{fig7:std_current_r}(c). However, as the flux phase increases to $\phi_M=5$ and $\phi_M=8$, additional crossings at non-zero voltage ($qV_b\neq 0$) emerge in the $G-V$ diagram [see Fig.~\ref{fig7:std_current_r}(i), (l)]. Remarkably, the number of non-zero crossing points in the $G-V$ characteristics equals to the crossing in the I-V hysteresis, despite occurring at different voltages [see the example of $\phi_M=5$ in Fig.~\ref{fig7:std_current_r}(h), (i), and also $\phi_M=8$ in Fig.~\ref{fig7:std_current_r}(k), (l)].
These results indicate that, at the condition of strongest resonance ($\mu=\epsilon_i'$), the number of crossing points grows systematically with the increase of $\phi_M$, providing a quantitative experimental signature of the active quantum memory. Thus, both the $\phi_M$ and the observed number of crossing points serve as descriptors of memory strength in an ion channel system. Since the number of crossing points can be directly measured from the I-V curves, our theory provides a practical experimental protocol for detecting quantum coherence effects in ion channel transport.

\begin{figure*}
	\centering
	\includegraphics[width=1\linewidth]{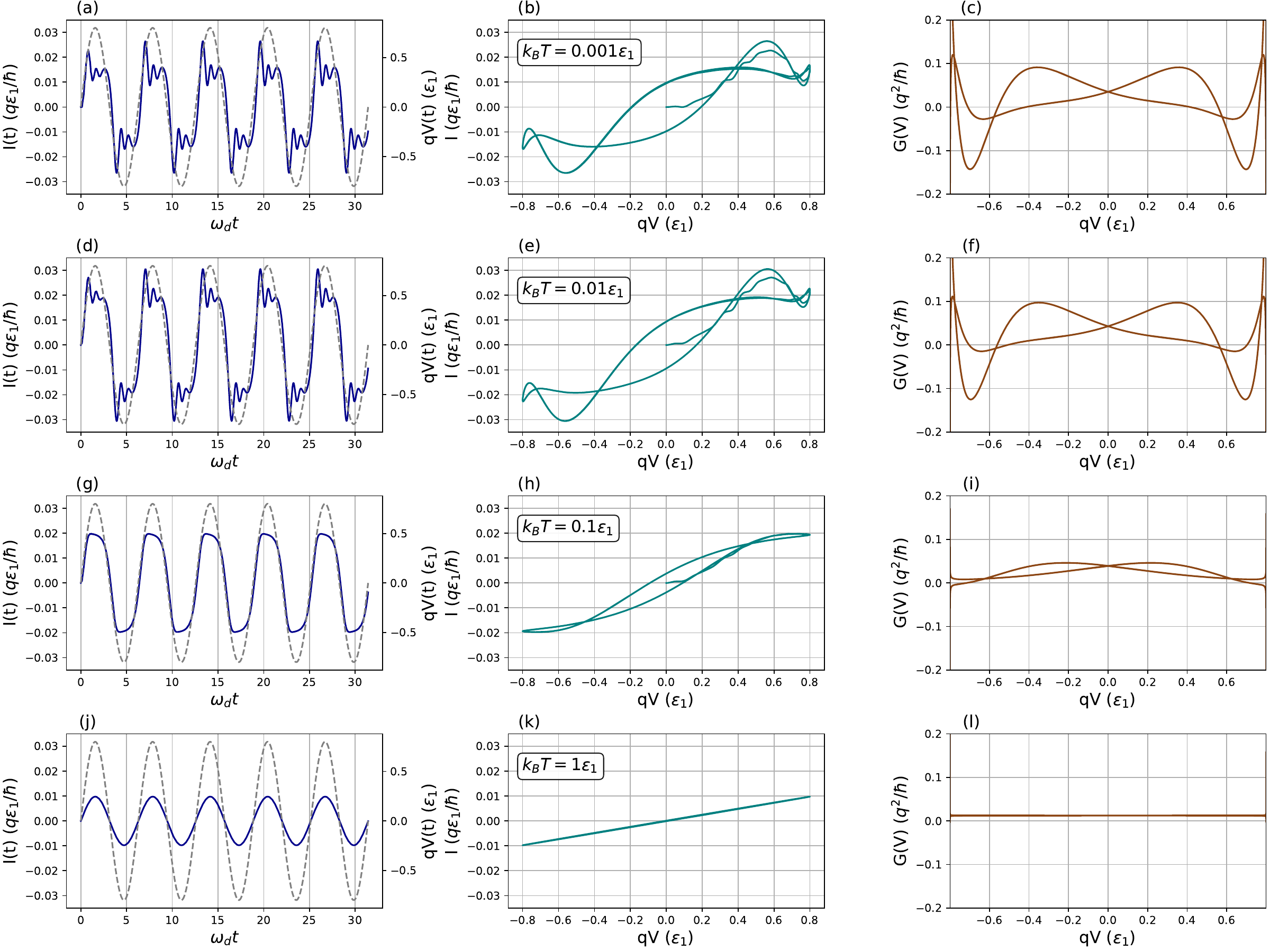}
	\caption{(color online). The temperature dependence of the ion channel current (dark-blue line) changed along with the sinusoidal bias voltage (dashed line), the I-V curve hysteresis (green line), and the steady-state differential conductance (brown line) under the oscillating bias voltage with different temperature (a)-(c) $k_BT = 0.001\epsilon_1$; (d)-(f) $k_BT = 0.01\epsilon_1$; (g)-(i) $k_BT = 0.1\epsilon_1$; and (j)-(l) $k_BT = 1\epsilon_1$, from the top panel to the bottom panel, respectively. The flux phase strength is fixed in $\phi_M = 8 $ with biasing voltage amplitude $qV_d=0.8\epsilon_1$. Other conditions are the same with Fig.~\ref{fig7:std_current_r}.}
	\label{fig8:std_current_T}
\end{figure*}

\subsection{The active quantum memory in the realistic ion channel energy scale at room temperature}
	
In the preceding analysis, the active quantum memory arises from the time-varying external membrane potential, and it is quantified by the magnetic flux phase strength $\phi_M$. Usually, at high temperatures, thermal fluctuation suppresses quantum coherence, causing the system to exhibit classical behavior. The suppression of quantum coherence causes the quantum memory to vanish as well. This decoherence phenomenon can be explained in terms of the two-time correlation memory kernel of Eq.~(\ref{eq:gt}) in our theory, by taking a high temperature ($k_BT\approx \epsilon_1$), compared to the system in a temperature much lower than $\epsilon_1$. Then the Fermi-Dirac distribution becomes nearly flat over the energy range in a constant number $f(\epsilon)\approx\frac{1}{2}$, and the memory kernel $\widetilde{g}^0_\alpha(t-t')$ is reduced to $\widetilde{g}^0_\alpha(t-t')\approx\delta(t-t')$. Correspondingly, the dynamics thus could become Markovian, namely memoryless. Consequently, at high temperatures, the system exhibits resistive behavior with linear I-V response (the Ohmic relation), eliminating all the quantum memory signatures. 

\begin{figure*}
	\centering
	\includegraphics[width=0.7\linewidth]{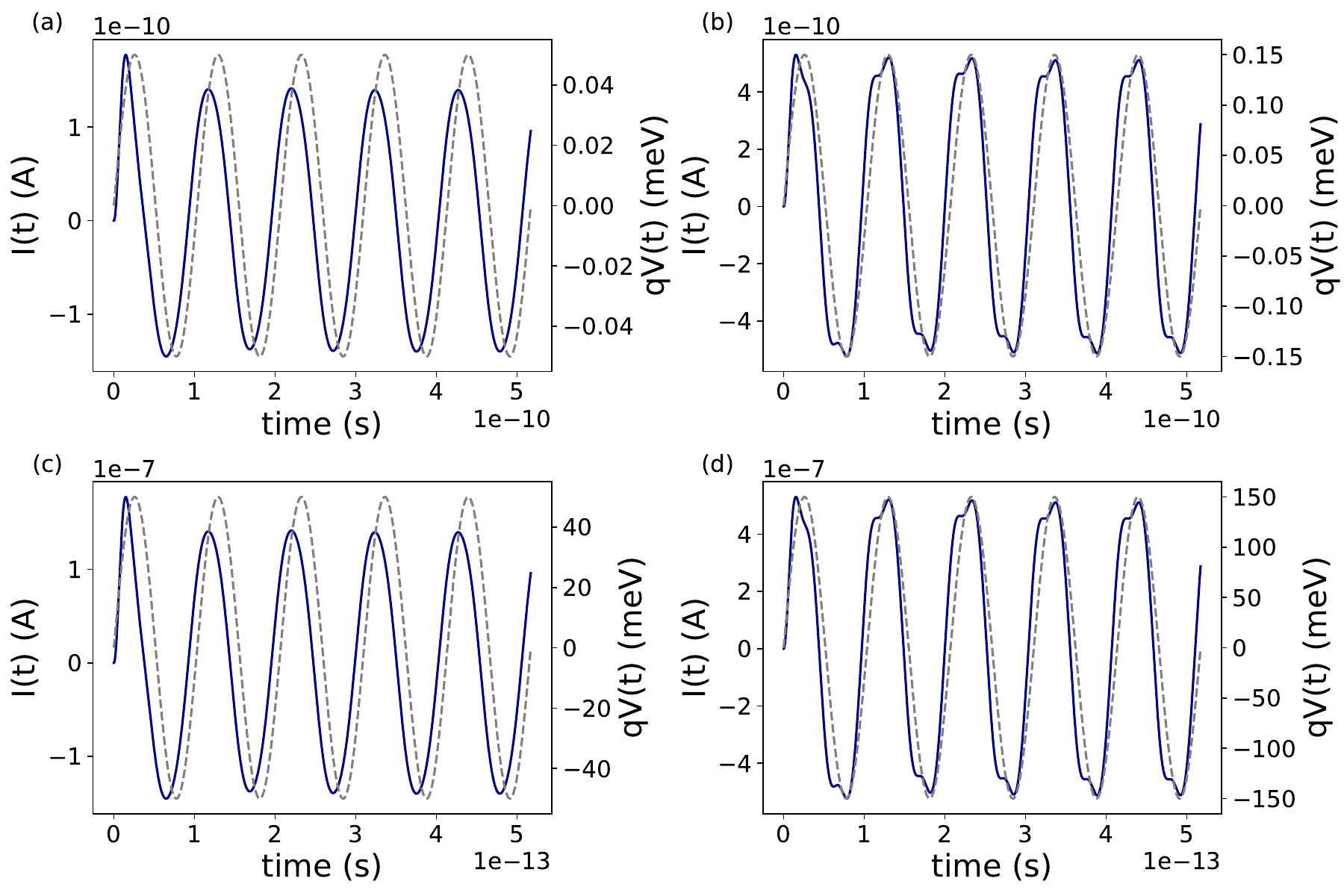}
	\caption{(color online). The comparison of the scaling unit between the electronic system and the neural system. (a) and (b) represents the scale of the electronic system at low temperature (0.29K). Two different voltage amplitudes are set (a) $V_{dot} = 0.05$ mV and (b) $V_{dot} = 0.15$ mV. (c) and (d) represents the scale of the ion channel system at room temperature (290 K). The voltage amplitudes are set (c) $V_d = 50$ mV and (d) $V_d = 150$ mV.}
	\label{fig9:scale}
\end{figure*}

To examine the temperature effect in our ion transport dynamics, we study the ion current, the corresponding I-V curve hysteresis, and the differential conductance by fixing $\phi_M=8$ across four different-scales of the temperatures: $kT=0.001\epsilon_1, 0.01\epsilon_1, 0.1\epsilon_1, 1\epsilon_1$. When the system is at low temperatures [$kT=0.001\epsilon_1$ and $0.01\epsilon_1$, Fig.~\ref{fig8:std_current_T}(a)-(f)], pronounced sidebands always present in the ion current. Both cases have four non-zero crossing points in the I-V hysteresis and G-V diagram. However, as the temperature increases to $kT=0.1\epsilon_1$ [see Fig.~\ref{fig8:std_current_T}(g)-(i)], the sidebands disappear, indicating the loss of the quantum coherence. Correspondingly, the number of crossing points decreases to two in both the I-V hysteresis and G-V diagram, confirming the loss of quantum memory. As the temperature further increases to $kT=1\epsilon_1$ [see Fig.~\ref{fig8:std_current_T}(j)-(l)], which is the fully Markovian regime, the ion current becomes pure sinusoidal form with the same phase as the driving voltage, and the sidebands disappear. The I-V relation becomes linear [see Fig.~\ref{fig8:std_current_T}(k)], and the conductance turns into a constant value [see Fig.~\ref{fig8:std_current_T}(l)], characterizing a resistive behavior obeying the classical Ohmic law. 
These results show that the thermal fluctuations at high temperatures can destroy quantum coherence. As thermal energy increases, the active quantum memory will decrease. This has challenged the existence of quantum effects in the room-temperature biological systems, even though some studies suggest that quantum coherence may persist in complex biological environments at room temperature \cite{MAR2018}. In the next, we will demonstrate the appearance of active quantum memory in a realistic ion channel system at room temperature.

Following the above analysis of quantum memory effects in ion channel transport currents under periodic time-dependent bias voltages, we now incorporate physiologically realistic parameters relevant to neural signal transmission. We show that active quantum memory effects can be observed experimentally through single ion channel recording techniques. The realistic ion channel energy scale is guided by the experimental data from patch clamp measurements. The Kv1.2 channel turns to be in the opening state from approximately $-$40 mV, and becomes totally opened at about $-$10 mV \cite{IRCI2015}. It remains in the opening state as the membrane voltage keeps increasing. Accordingly, we model the transmembrane potential as a periodic voltage $V_b(t)$ varying in a range from $-$150 mV to 150 mV. Based on characteristic energy barriers for ion transport of approximately $2-4$ kcal/mol (approximately $87 \sim 174$ meV)\cite{BR2001}, we set the on-site energies as $\epsilon_1 = \epsilon_4 = 120$ meV and $\epsilon_2 = \epsilon_3 = 180$ meV, reflecting a symmetric potential profile within the selectivity filter.  The hopping energy is set to be $J=30$ meV for all adjacent sites. The coupling strengths between the channel binding sites and the extracellular and intracellular electrolytic solutions are characterized by coupling constants $\Gamma_E = \Gamma_I = 10$ meV. The chemical potentials of both sides are the same as $\mu_E = \mu_I = 10$ meV, denoting that the membrane potential difference is mainly dominated by $V_b(t)$. The temporal dynamics of the ion channel current are primarily governed by the time-dependent voltage $V_b(t)$, with a driving frequency corresponding to $\hbar\omega_d = 40$ meV, consistent with the characteristic energy scale of the system.

\begin{figure*}
	\centering
	\includegraphics[width=0.9\linewidth]{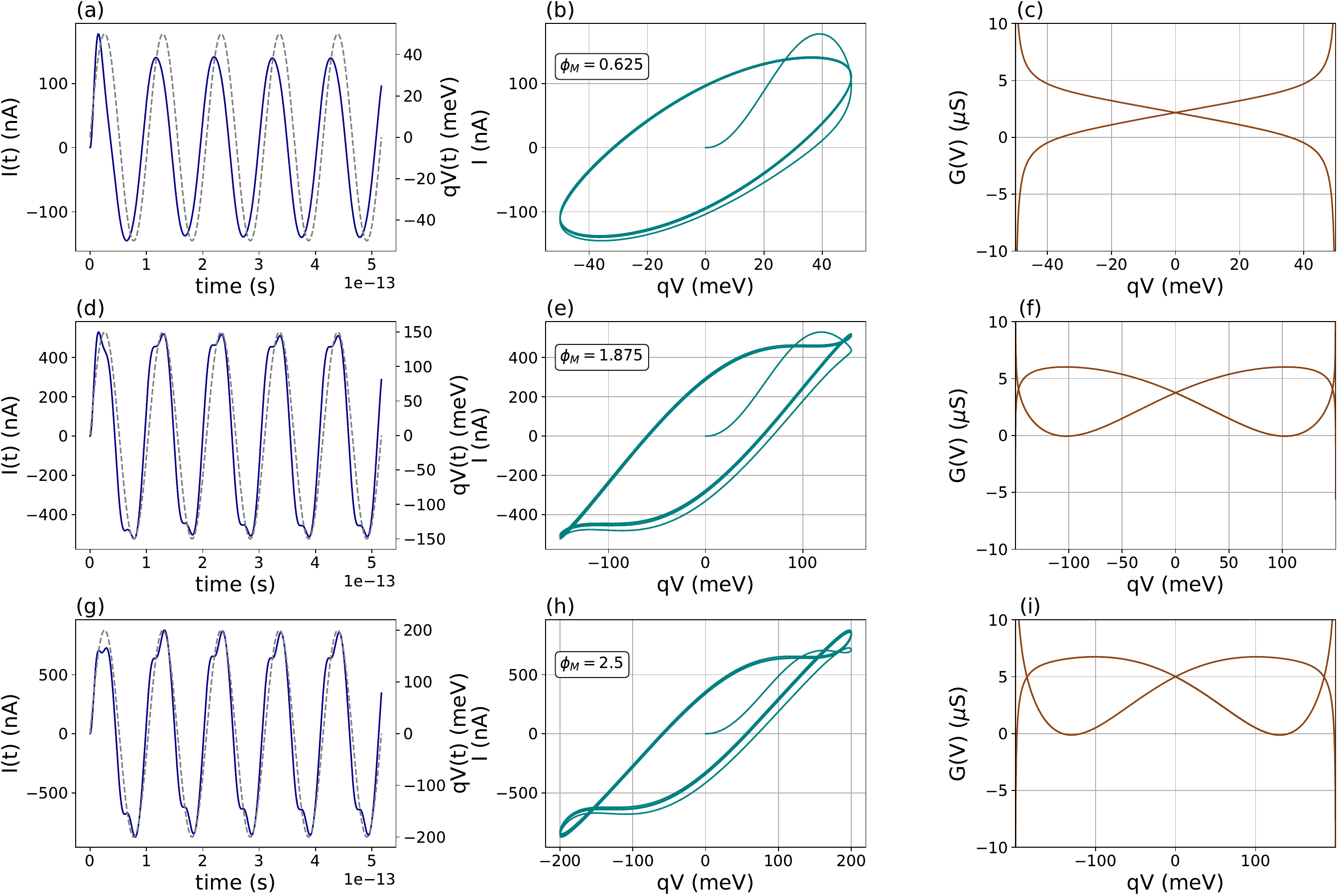}
	\caption{(color online). The ion transport currents, the I-V characteristics and the differential conductances with different voltage amplitude in the neural scaling unit, which are set as (a)-(c) $qV_d = 50$ meV; (d)-(f) $qV_d = 150$ meV; (g)-(i) $qV_d = 200$ meV with oscillating frequency $\hbar\omega_d = 40$ meV (about 9.68 THz). All of them are set in the room temperature 290 K. Other parameters are the same with Fig.~\ref{fig9:scale}(c) and (d).
	}
	\label{fig10:ion channel}
\end{figure*}

The typical temperature in physiological neuronal environments is approximately 20 °C, corresponding to a thermal energy of $k_B T \approx 25$ meV $\approx 0.2 \epsilon_1$. This shows that thermal fluctuations are not large enough to destroy the ion coherent tunnelings. Given that energy scales and temperatures in nanoelectronic systems, where quantum coherent tunnelings are dominant in the electron transport, can be proportionally scaled to match those in ion channel systems, we find that ion channels operating at room temperature can have quantum memory effects. To illustrate this correspondence explicitly, we consider an analogous multi-level quantum dot device in which all energy parameters are scaled down by a factor of 1000 to 10000 relative to the ion channel system due to the mass scale difference between ions and electrons. Thus, we take the quantum dot energy levels with $\epsilon_1 = \epsilon_4 = 0.12$ meV and $\epsilon_2 = \epsilon_3 = 0.18$ meV, with nearest-neighbor tunneling energy $J = 0.03$ meV. The coupling strengths to the source and drain leads are $\Gamma_S = \Gamma_D = 0.01$ meV. The operating temperature is about 0.29 K, corresponding to a thermal energy of $k_B T = 0.025$ meV. The chemical potentials of both leads are set equal, $\mu_S = \mu_D = 0.01$ meV, so that the current is driven solely by the time-dependent bias voltage, which oscillates at $\hbar\omega_{\text{dot}} = 0.04$ meV (about $10^{11}$ Hz). We also apply sinusoidal voltage modulations for five complete cycles, with two amplitudes, $eV_{\text{dot}} = 0.05$ meV and $eV_{\text{dot}} = 0.15$ meV. The resulting electronic transport currents are shown in Fig.~\ref{fig9:scale}(a) and Fig.~\ref{fig9:scale}(b), respectively.

Then, for the ion channel system, we apply periodic voltage modulations with two amplitudes, $qV_d = 50$ meV and $qV_d = 150$ meV, where $q$ is the charge of a single ion. The driving frequency is set as $\hbar\omega_d = 40$ meV, corresponding to about $f_d = 9.68 \times 10^{12}$ Hz. The currents are shown in Fig.~\ref{fig9:scale}(c) and (d). The close correspondence between Fig.~\ref{fig9:scale}(a) and (c), as well as between Fig.~\ref{fig9:scale}(b) and (d), demonstrates that these two physically distinct systems, electrons in quantum dots at cryogenic temperatures and ions in biological channels at room temperature, show the equivalent quantum transport characteristics under appropriately scaled parameters. The sideband current features observed in Fig.~\ref{fig9:scale}(d) reveal that even at room temperature, active quantum memory effects persist when both the magnetic flux phase strength $\phi_M$ and the voltage amplitude are sufficiently large to overcome thermal fluctuations. Building on these observations, we analyze the signatures of active quantum memory in the ion channel system using the same approach as in the previous section, focusing on the crossing points in the current-voltage characteristics and the features of the differential conductance.

In Fig.~\ref{fig10:ion channel}, we examine the ion channel system at room temperature (approximately 20°C) under varying voltage amplitudes ($qV_d = 50$ meV, 150 meV, and 200 meV) with a fixed driving frequency of $\hbar\omega_d = 40$ meV. The first two cases ($qV_d = 50$ meV and 150 meV) employ the same parameters as those in Fig.~\ref{fig9:scale}(c) and (d). At $qV_d = 50$ meV, corresponding to a lower phase strength of $\phi_M = 0.625$, the current-voltage characteristic exhibits no crossing points [see Fig.~\ref{fig10:ion channel}(b)], and the differential conductance displays no zero-crossing features [see Fig.~\ref{fig10:ion channel}(c)], where the quantum memory is suppressed. However, when the voltage increases to $qV_d = 150$ meV with the phase strength $\phi_M = 1.875$ while maintaining the same frequency, the current develops sideband structures and the I-V characteristic slightly shows two crossing points, as shown in Fig.~\ref{fig10:ion channel}(d)-(f). In addition, the zero-crossing feature is more pronounced in the G-V diagram. These observations are consistent with the predictions presented in Fig.~\ref{fig7:std_current_r}. Upon further increasing the voltage amplitude to 200 meV, as depicted in Fig.~\ref{fig10:ion channel}(g)-(i), the two crossing points in the I-V characteristics become more pronounced, further corroborating the presence of active quantum memory effects at room temperature.

In summary, these results demonstrate that single ion channels can exhibit active quantum memory under THz-frequency time-varying potentials, even at room temperature. The quantum coherence induced by the external voltage at the characteristic energy scale of the ion channel manifests through the sideband behaviour in the ion steady-state current, the non-zero crossing points of I-V characteristics, and the corresponding non-zero crossing points of G-V diagrams. 
We note that the current magnitudes in Fig.~\ref{fig9:scale} and Fig.~\ref{fig10:ion channel} are approximately several hundred 
nanoamperes, about three orders of magnitude larger than the picoampere-level 
currents typically observed in patch-clamp experiments.
This discrepancy arises because patch-clamp measurements measure not an isolated single ion channel, but rather the average of a complex circuit 
composed of numerous ion channels embedded in the cell membrane.
The cell membrane is composed of a lipid bilayer, which functions similarly to a capacitor group in parallel among the ion channels. 
A large current is required to charge or discharge the membrane capacitance. These capacitive transient changes can distort or completely 
mask the rapid opening and closing of single ion channels, thereby could 
effectively reduce the applied high-frequency voltage.  Also, the resistance of the pipette tip acts in series with the 
membrane, causing the actual voltage experienced by the membrane is lower than the applied voltage. 
Nevertheless, our single-channel model preserves the sideband structure associated with active quantum memory and reveals quantum coherence
under the applied high-frequency voltage. These theoretical predictions suggest that experimental 
observation of active quantum memory effect in ion channels requires an isolated single ion channel measurement with high temporal resolutions.

\section{Discussions and Perspectives}
In this work, we construct a quantum ion channel model for the ion selectivity filter region to describe the ion channel dynamics. We model the selectivity filter as a multi-level ionic tunneling junction and study the ionic quantum transport. By deriving the quantum Langevin equation from the first principle of quantum mechanics and the ion transport current in terms of the non-equilibrium Green's functions, we can study the ion channel dynamics in the open quantum system framework.  We find that the external time-varying voltage applied across the cell membrane induces an effective magnetic flux phase. It further causes quantum coherence to emerge in ion tunneling and forms quantum memory in the ion system. We define this memory dynamics as the active quantum memory, and show that the active quantum memory is manifested in terms of the sideband oscillations of the ionic current. We also find that the active quantum memory can be depicted by the non-zero crossing points in both the I-V hysteresis and the G-V diagram, which provides a quantitative way for observing the active quantum memory in experiments. This also gives an underlying quantum description of the active quantum memory in the ion channel under neuronal membrane stimulation.

Furthermore, we investigate the extent of active quantum memory for different temperatures. Specifically, we examine whether quantum coherence disappears at high temperatures. As expected, the number of crossing points in both the I-V hysteresis and the G-V diagram decreases when the temperature increases significantly. When the temperature becomes very high relative to $\epsilon_1$, the system retains no memory, and the I-V relation reduces to an Ohmic relation. Given that the ion channels operate at room temperature, we also demonstrate the results on the real biological scale. By estimating the characteristic energy scale of the Kv1.2 channel using the free energy barrier and other relevant data, we find that the ion dynamics in the ion channel exhibit active quantum memory at room temperature. This result demonstrates that under physiological conditions with appropriate external stimulation, the active quantum memory can be observed in the ion channel dynamics.

However, experimental verification of these predictions faces significant technical challenges. To observe the sideband oscillation behaviour in the ion current, such as non-zero crossing points in I-V hysteresis and differential conductance calculated in this work, the temporal resolution of measuring single ion channel current must be improved from milliseconds to nanoseconds and even picoseconds. The implementation of these results would require a terahertz-frequency oscillating bias voltage in the patch clamp experiments and the nanosecond to picosecond time resolution in the current measurement, while such high-frequency protocols have not yet been implemented. Practically, due to limitations in signal-to-noise ratio affecting the sampling frequency, the time resolution of single-ion signal recordings remains lower, mostly at the millisecond resolution in most of the electrophysiology experiments \cite{SN1983, SP, MS}. It is worth mentioning that a cross-domain and novel method using the micro- or nano-electronic structure measuring the single ion channel signal supports a higher bandwidth to about 500kHz-10MHz \cite{FKGB2002, RRRS2013, HOS2018, SCH2019}, corresponding to the time resolution of microseconds. Except for the nanostructure measurement, there are optical stimulation and all-optical electrophysiological techniques, providing high-speed and non-invasive measurements \cite{HZFK2014, ZRC2016, FC2021}, and the two-dimensional infrared spectra, which capture the ultrafast snapshot of ions, providing an instant ion configuration in the selectivity filter with the time resolution in picoseconds \cite{KCM2016}. Our theoretical predictions provide a concrete experimental protocol for detecting quantum coherence in ion channels. We believe that such results would be achievable with the advanced techniques in the future.

To utilize our quantum mechanical approach within the broader landscape of ion channel research, we note that current experimental observations of ion channel dynamics typically focus on collective behavior in channel clusters. The advances in structural biology, including cryo-EM and high-resolution NMR, have enabled the detailed structural analysis of ion channels in various functional states. Complementarily, optogenetics \cite{TF2021, GGS2022} and chemogenetics \cite{SR2014, Roth2016, CM2018} techniques enable precise manipulation of ion channels in living cells. Additionally, the ion transports in the channel under the THz resonant field have been analyzed theoretically \cite{LCZ2021, HLH2022, LLS2024, SXO2025} and experimentally \cite{LQC2021}, supporting the evidence that the resonance with the selectivity filter enhances the permeation of ions in the channel. These techniques have improved the understanding of the dynamic role of ion channels in the neural system. We hope that this work, based on the underlying quantum dynamics, can provide a starting point for a quantum mechanics-based description of ion channel dynamics.

In this study, we only model and analyze single ion channel dynamics as a simple solvable system in quantum mechanical theory, which is still far from the complexity of real neural system. Nonetheless, our findings raise intriguing questions about potential connection to higher-level psychological functions. A specific memory concept related to learning and cognition is the so-called working memory \cite{MGP1960, AS1968, BH1974, Cow1988, Cow1999, Ba2000}. The dependence of active quantum memory on the amplitude and the frequency of oscillating voltage shares similar characteristics with working memory in cognitive neuroscience. Specifically, both the active quantum memory and the working memory are enhanced when the external stimuli have a higher amplitude or a slower and more sustainable duration. Based on our theory, we anticipate that working memory may be a collective phenomenon of active quantum memory. This gives a possible explanation for the physical principle of working memory at the microscopic level. 

While such parallels are suggestive, we emphasize that establishing a direct mechanistic link between single-channel quantum coherence and macroscopic cognitive phenomena remains highly speculative at present. Given that the channel potential changes measured by both the voltage clamp (such as the action potential in Hodgkin and Huxley model) or the patch clamp in ion channel experiments are around $100-200$ mV, indicating that the temperature effect is not the dominant factor. Thus, we speculate that the quantum memory effect will exist in more than one ion channel, where the Hamiltonian should be more complex. Our present work focuses on the fundamental quantum transport of ion channel selectivity filter and provides a starting point for investigations into whether quantum effects at the molecular level can contribute to neural dynamics. Such explorations, while challenging, could offer new perspectives on the physical foundations of biological systems.

\ \\
\acknowledgments

This work was supported by National Science and Technology Council of Taiwan, Republic of China, under Contract No. NSTC-111-2112-M-006-014-MY3.


\begin{references}
	\bibitem{HH1952} A. L. Hodgkin, and A. F. Huxley, \textit{A quantitative description of membrane current and its application to conduction and excitation in nerve}, J. Physiol. \textbf{117}, 500 (1952).

	\bibitem{NS1976} E. Neher, and B. Sakmann, \textit{Single-channel currents recorded from membrane of denervated frog muscle fibres}, Nature \textbf{260}, 799 (1976).

	\bibitem{HMNSS1981} O. P. Hamill, A. Marty, E. Neher, B. Sakmann and F. J. Sigworth, \textit{Improved patch-clamp techniques for high-resolution current recording from cells and cell-free membrane patches}, Pflugers Arch. \textbf{391}, 85 (1981).

	\bibitem{Do1998} D. A. Doyle, J. Morais Cabral, R. A. Pfuetzner, A. Kuo, J. M. Gulbis, S. L. Cohen, B. T. Chait, and R. MacKinnon, \textit{The Structure of the Potassium Channel: Molecular Basis of $K^+$ Conduction and Selectivity}, Science \textbf{280}, 69 (1998).

	\bibitem{Mac2003} R. MacKinnon, \textit{Potassium channels}, FEBS Lett. \textbf{555}, 62 (2003).
	
	\bibitem{Ca2010} W. A. Catterall, \textit{Ion channel voltage sensors: structure, function, and pathophysiology}, Neuron \textbf{67}, 915 (2010).

	\bibitem{SMSS2018} A. K. Singh, L. L. McGoldrick, K. Saotome, and A. I. Sobolevsky, \textit{X-ray crystallography of TRP channels}, Channels \textbf{12}, 137 (2018).

	\bibitem{GT2020} J. García-Nafría, C. G. Tate. , \textit{Cryo-Electron Microscopy: Moving Beyond X-Ray Crystal Structures for Drug Receptors and Drug Development}, Annu. Rev. Pharmacol. Toxicol. \textbf{60}, 51 (2020).

	\bibitem{MZG2021} A. Mironenko, U. Zachariae, B. L. de Groot, and W. Kopec, \textit{The persistent question of potassium channel permeation mechanisms}, J. Mol. Biol. \textbf{433}, 167002 (2021).
	
	\bibitem{WYYBR2025} Y. Wu, Y. Yan, Y. Yang, S. Bian, A. Rivetta, K. Allen, F. J Sigworth  \textit{CryoEM structures of Kv1.2 potassium channels, conducting and non-conducting}, eLife \textbf{12}, RP89459 (2025).
	
	
	\bibitem{AH1998} C. M. Armstrong, and B. Hille, \textit{Voltage-gated ion channels and electrical excitability}, Neuron \textbf{20}, 371 (1998).
	
	\bibitem{NMPD2016} D. Naranjo, H. Moldenhauer, M. Pincuntureo, I. Díaz-Franulic, \textit{Pore size matters for potassium channel conductance}, J. Gen. Physiol. \textbf{148}, 277 (2016).
	
	\bibitem{KRST2019} G. Klesse, S. Rao, M. S. P. Sansom, and S. J. Tucker, \textit{CHAP: A Versatile Tool for the Structural and Functional Annotation of Ion Channel Pores}, J. Mol. Biol. \textbf{431}, 3353 (2019).
		
	\bibitem{CMDM1997} C. C. Cruickshank, R. F. Minchin, A. C. Le Dain, and B. Martinac, \textit{Estimation of the pore size of the large-conductance mechanosensitive ion channel of Escherichia coli}, Biophys. J. \textbf{73}, 1925 (1997).
	
	\bibitem{BR2001} S. Bernèche, B. Roux, \textit{Energetics of ion conduction through the $K^+$ channel}, Nature \textbf{414}, 73 (2001).
	
	\bibitem{MZM2001} J. H. Morais-Cabral, Y. Zhou, and R. MacKinnon, \textit{Energetic optimization of ion conduction rate by the $K^+$ selectivity filter}, Nature \textbf{414}, 37 (2001).
		
	\bibitem{ZMCKM2001} Y. Zhou, J. H. Morais-Cabral, A. Kaufman, R. MacKinnon, \textit{Chemistry of ion coordination and hydration revealed by a $K^+$ channel-Fab complex at 2.0 \AA ~resolution}, Nature \textbf{414}, 43 (2001).
	
	\bibitem{VP2010} A. Vaziri and M. B. Plenio, \textit{Quantum coherence in ion channels: resonances, transport and verification}, New J. Phys. \textbf{12}, 085001 (2010).

	\bibitem{HHCSC2010} L.T. Hall, C.D. Hill, J.H. Cole, B. Städler, F. Caruso, P. Mulvaney, J. Wrachtrup, and L.C.L. Hollenberg, \textit{Monitoring ion-channel function in real time through quantum decoherence}, Proc. Natl. Acad. Sci. U.S.A. \textbf{107}, 18777 (2010).
	
	\bibitem{SSB2012}J.~Summhammer, V.~Salari, and G.~Bernroider, \textit{A quantum-mechanical description of ion motion within the confining potentials of voltage-gated ion channels}, J. Integr. Neurosci. \textbf{11}, 123 (2012).
		
	\bibitem{SNS2017} V.~Salari, H.~Naeij, and A.~Shafiee,\textit{Quantum Interference and Selectivity through Biological Ion Channels}, Sci. Rep. \textbf{7}, 41625 (2017).
	
	\bibitem{RABI} B. Roux, T. Allen, S. Bernèche, W. Im, \textit{Theoretical and computational models of biological ion channels}, Q. Rev. Biophys. \textbf{37}, 15 (2004).
	
	\bibitem{MBYWA2012} C. Maffeo, S. Bhattacharya, J. Yoo, D. Wells, and A. Aksimentiev, \textit{Modeling and simulation of ion channels}, Chem. Rev. \textbf{112}, 6250 (2012).
	
	\bibitem{MWK} N. Modi, M. Winterhalter, and U. Kleinekathöfer, \textit{Computational modeling of ion transport through nanopores}, Nanoscale textbf{4}, 6166 (2012).
	
	\bibitem{HJ1996} H. Haug and A.-P. Jauho, \textit{Quantum Kinetics in Transport and Optics of Semiconductors} (Springer, Berlin,
	1996).
	
	\bibitem{LCM} S. B. Long, E. B. Campbell, and R. MacKinnon, \textit{Crystal Structure of a Mammalian Voltage-Dependent Shaker Family $K^+$ Channel}, Science \textbf{309}, 897 (2005).
	
	\bibitem{KPH} Q. Kuang, P. Purhonen, and H. Hebert, \textit{Structure of potassium channels}, Cell. Mol. Life Sci. \textbf{72}, 3677 (2015).
	
	\bibitem{KME} I. Kaufman, P. V. E. McClintock and R. S. Eisenberg, \textit{Coulomb blockade model of permeation and selectivity in biological ion channels}, New J. Phys. \textbf{17}, 083021 (2015).
	
	\bibitem{TuZhang2008} M. W. Y. Tu and W.-M. Zhang, \emph{Non-markovian decoherence theory for a double-dot charge qubit}, Phys. Rev. B \textbf{78}, 235311 (2008).
		
	\bibitem{JTZY} J.~Jin, M. W.-Y. Tu, W.-M. Zhang, and Y. Yan, \textit{Non-equilibrium quantum theory for nanodevices based on the Feynman-Vernon influence functional}, New J. Phys. \textbf{12}, 083013 (2010).
		
	\bibitem{LZ2012} C. U. Lei and W.-M. Zhang, \emph{A quantum photonic dissipative transport theory}, Ann. Phys. \textbf{327}, 1408 (2012).
	
	\bibitem{YLZ} P.-Y.~Yang, C.-Y.~Lin, and W.-M.~Zhang, \textit{Master equation approach to transient quantum transport in nanostructures incorporating initial correlations}, Phys. Rev. B \textbf{92}, 165403 (2015).
	
	\bibitem{YZ2017} P.-Y.~Yang and W.-M.~Zhang, \emph{Master equation approach to transient quantum transport in nanostructures}, Front. Phys. \textbf{12}, 127204 (2017).
	
	\bibitem{YLZ2023} C.-Z.~Yao, H.-L.~Lai, and W.-M.~Zhang, \emph{Quantum transport theory of hybrid superconducting systems}, Phys. Rev. B \textbf{108}, 195402 (2023).

	\bibitem{ZLXTN} W.-M. Zhang, P.Y. Lo, H.-N. Xiong, M. W.-Y. Tu and F. Nori, \textit{General Non-Markovian Dynamics of Open Quantum Systems}, Phys. Rev. Lett. \textbf{109}, 170402 (2012).
	
	\bibitem{WMZ2019} W.-M. Zhang, \emph{Exact master equation and general non-Markovian dynamics in open quantum systems}, Eur. Phys. J. Special Topics, \textbf{227}, 1849 (2019).
		
	\bibitem{WJM} N. S. Wingreen, A.-P. Jauho, and Y. Meir, \textit{Time-dependent transport through a mesoscopic structure}, Phys. Rev. B \textbf{48}, 8487(R) (1993).
	
	\bibitem{JWM} A.-P. Jauho, N. S. Wingreen, and Y. Meir, \textit{Time-dependent transport in interacting and noninteracting resonant-tunneling systems}, Phys. Rev. B \textbf{50}, 5528 (1994).	
	
	\bibitem{Tannoudji2020} C. Cohen-Tannoudji, B. Diu, and F. Laloë, \emph{Quantum mechanics: Vol 1: Basic concepts, tools, and applications} (Wiley-VCH, Weinheim, 2020).
	
	\bibitem{MAR2018} A. Marais, B. Adams, A. K. Ringsmuth, M. Ferretti, J. M. Gruber, R. Hendrikx, M. Schuld, S. L. Smith, I. Sinayskiy, T. P. J. Krüger, F. Petruccione, R. v. Grondelle, \textit{The future of quantum biology}, J R Soc Interface \textbf{15}, 20180640 (2018).
	
	\bibitem{IRCI2015} I. G. Ishida, G. E. Rangel-Yescas, J. Carrasco-Zanini, L. D. Islas, \textit{Voltage-dependent gating and gating charge measurements in the Kv1.2 potassium channel}, J Gen Physiol \textbf{145}, 345 (2015).
	
	\bibitem{SN1983} B. Sakmann and E. Neher, \textit{Single- Channel Recording} (Plenum, New York, 1983)
	
	\bibitem{SP} J. Shuai, and I. Parker, \textit{Optical single-channel recording by imaging $Ca^{2+}$ flux through individual ion channels: theoretical considerations and limits to resolution}, Cell calcium \textbf{37}, 283 (2005).
	
	\bibitem{MS} M. Mortensen, and T. G. Smart, \textit{Single-channel recording of ligand-gated ion channels}, Nat. Protoc. \textbf{2}, 2826 (2007).
	
	\bibitem{FKGB2002} N. Fertig, M. Klau, M. George, R. H. Blick, and J. C. Behrends, \textit{Activity of single ion channel proteins detected with a planar microstructure}, Appl. Phys. Lett. \textbf{81}, 4865 (2002).
	
	\bibitem{RRRS2013} J. K. Rosenstein, S. Ramakrishnan, J. Roseman, and K. L. Shepard, \textit{Single ion channel recordings with CMOS-anchored lipid membranes}, Nano Lett. \textbf{13}, 2682 (2013).
	
	\bibitem{HOS2018} A. J. W. Hartel, P. Ong, I. Schroeder, M. H. Giese, S. Shekar, O. B. Clarke, R. Zalk, A. R. Marks, W. A. Hendrickson, and K. L. Shepard, \textit{Single-channel recordings of RyR1 at microsecond resolution in CMOS-suspended membranes}, Proc. Natl. Acad. Sci. U.S.A.\textbf{115}, E1789 (2018).
	
	\bibitem{SCH2019} S. Shekar, C.-C. Chien, A. Hartel, P. Ong, O. B. Clarke, A. Marks, M. Drndic, and K. L. Shepard, \textit{Wavelet Denoising of High-Bandwidth Nanopore and Ion-Channel Signals}, Nano Letters \textbf{19}, 1090 (2019).
	
	\bibitem{HZFK2014} D. R. Hochbaum, Y. Zhao, S. L. Farhi, N. Klapoetke, C. A. Werley, V. Kapoor, P. Zou, J. M. Kralj, D. Maclaurin, N. Smedemark-Margulies, et al., \textit{All-optical electrophysiology in mammalian neurons using engineered microbial rhodopsins}, Nat. Methods \textbf{11}, 825 (2014).
	
	\bibitem{ZRC2016} H. Zhang, E. Reichert, A. E. Cohen, \textit{Optical electrophysiology for probing function and pharmacology of voltage-gated ion channels}, eLife \textbf{5}, 15202 (2016).
	
	\bibitem{FC2021} L. Filipis, and M. Canepari, \textit{Optical measurement of physiological sodium currents in the axon initial segment}, J. Physiol. \textbf{599}, 49 (2021).
	
	\bibitem{KCM2016} H. T. Kratochvil, J. K. Carr, K. Matulef, A. W. Annen, H. Li, M. Maj, J. Ostmeyer, A. L. Serrano, H. Raghuraman, S. D. Moran et al., \textit{Instantaneous ion configurations in the $K^+$ ion channel selectivity filter revealed by 2D IR spectroscopy}, science \textbf{353}, 1040 (2016).
	
	\bibitem{TF2021} H. Tsukamoto, Y. Furutani, in \textit{Optogenetics}, \textit{Optogenetic Modulation of Ion Channels by Photoreceptive Proteins}, edited by H. Yawo, H. Kandori, A. Koizumi, and R. Kageyama (Springer, Singapore, 2021)
	
	\bibitem{GGS2022} E. G. Govorunova, Y. Gou, O. A. Sineshchekov, H. Li, X. Lu, Y. Wang, L. S. Brown, F. St-Pierre, M. Xue, and J. L. Spudich, \textit{Kalium channelrhodopsins are natural light-gated potassium channels that mediate optogenetic inhibition}, Nat. Neurosci. \textbf{25}, 967 (2022). 
	
	\bibitem{SR2014} S. M. Sternson, and B. L. Roth, \textit{Chemogenetic Tools to Interrogate Brain Functions}, Annu. Rev. Neurosci. \textbf{37}, 387 (2014).
	
	\bibitem{Roth2016} B. L. Roth, \textit{DREADDs for Neuroscientists}, Neuron \textbf{89}, 683 (2016).
	
	\bibitem{CM2018} E. J. Campbell, N. J. Marchant, \textit{The use of chemogenetics in behavioural neuroscience: receptor variants}, Br. J. Pharmacol. \textbf{175}, 994 (2018).
	
	\bibitem{LCZ2021} Y. Li, C. Chang, Z. Zhu, L. Sun, and C. Fan, \textit{Terahertz Wave Enhances Permeability of the Voltage-Gated Calcium Channel}, J. Am. Chem. Soc. \textbf{143}, 4311 (2021).
	
	\bibitem{HLH2022} Z.-H. Hu, W.-P. Lv, D.-X. Hui, X.-J. Wang, and Y.-N. Wang, \textit{Permeability enhancement of the KcsA channel under radiation of a terahertz wave}, Phys. Rev. E \textbf{105}, 024104 (2022).
	
	\bibitem{LLS2024} Y. Liu, X. Liu, Y. Shu, and Y. Yu, \textit{Progress of the Impact of Terahertz Radiation on Ion Channel Kinetics in Neuronal Cells}, Neurosci Bull \textbf{40}, 1960 (2024).
	
	\bibitem{SXO2025} Z. Song, L. Xue, Q. Ouyang, and C. Song, \textit{Impact of a Terahertz electromagnetic field on the ion permeation of potassium and sodium channels}, Commun Chem \textbf{8}, 101 (2025). 
	
	\bibitem{LQC2021} X. Liu, Z. Qiao, Y. Chai, Z. Zhu, K. Wu, W. Ji, D. Li, Y. Xiao, L. Mao, C. Chang et al., \textit{Nonthermal and reversible control of neuronal signaling and behavior by midinfrared stimulation}, Proc. Natl. Acad. Sci. U. S. A. \textbf{118}, e2015685118 (2021). 
	
	\bibitem{MGP1960} G. A. Miller, E. Galanter, and K. H. Pribram, \textit{Plans and the Structure of Behaviour} (Henry Holt and Co., New York, 1960).
	
	\bibitem{AS1968} R. C. Atkinson, R. M. Shiffrin, in \textit{The psychology of learning and motivation: Advances in research and theory, Human memory: A proposed system and its control processes}, edited by K. W. Spence and J. T. Spence (Academic Press, New York, 1968), Vol. 2, pp. 89-195.
	
	\bibitem{BH1974} A. D. Baddeley, and G. Hitch, in \textit{The psychology of learning and motivation: Advances in research and theory, Working memory}, edited by G. H. Bower (Academic Press, New York, 1974), Vol. 8, pp. 47–89.
	
	\bibitem{Cow1988} N. Cowan, \textit{Evolving conceptions of memory storage, selective attention, and their mutual constraints within the human information-processing system}, Psychol. Bull. \textbf{104}, 163 (1988).
	
	\bibitem{Cow1999} N. Cowan, in \textit{Models of Working Memory: Mechanisms of Active Maintenance and Executive Control, An Embedded-Processes Model of Working Memory}, edited by A. Miyake and P. Shah (Cambridge University Press, Cambridge, 1999), pp. 62-101.
	
	\bibitem{Ba2000} A. Baddeley, \textit{The episodic buffer: a new component of working memory?}, Trends Cogn. Sci. \textbf{4}, 417 (2000).	
\end{references}
\end{document}